\documentclass[fleqn,usenatbib]{mnras}

\usepackage{newtxtext,newtxmath}
\usepackage{mathtools}

\usepackage[T1]{fontenc}

\usepackage{graphicx}
\usepackage{amsmath}
\usepackage{soul}
\usepackage[normalem]{ulem}
\usepackage{xspace}
\usepackage[dvipsnames]{xcolor}
\usepackage{pifont}
\usepackage{tabularx}
\usepackage{dirtree}
\usepackage{orcidlink}
\usepackage{makecell}

\newcommand{\cmark}{\textcolor{ForestGreen}{\ding{51}}}%
\newcommand{\xmark}{\textcolor{BrickRed}{\ding{55}}}%

\usepackage{multirow}

\makeatletter 
  
  \patchcmd{\NAT@citex}
    {\@citea\NAT@hyper@{%
      \NAT@nmfmt{\NAT@nm}%
      \hyper@natlinkbreak{\NAT@aysep\NAT@spacechar}{\@citeb\@extra@b@citeb}%
      \NAT@date}}
    {\@citea\NAT@nmfmt{\NAT@nm}%
    \NAT@aysep\NAT@spacechar\NAT@hyper@{\NAT@date}}{}{}

  \patchcmd{\NAT@citex}
    {\@citea\NAT@hyper@{%
      \NAT@nmfmt{\NAT@nm}%
      \hyper@natlinkbreak{\NAT@spacechar\NAT@@open\if*#1*\else#1\NAT@spacechar\fi}%
        {\@citeb\@extra@b@citeb}%
      \NAT@date}}
    {\@citea\NAT@nmfmt{\NAT@nm}%
    \NAT@spacechar\NAT@@open\if*#1*\else#1\NAT@spacechar\fi\NAT@hyper@{\NAT@date}}
    {}{}
\makeatother

\newcommand{\LCDM}{$\Lambda$CDM\xspace}

\newcommand{\gadgetfour}{\textsc{GADGET-4}\xspace}

\newcommand{\Msun}{{\rm M_\odot}}

\newcommand{\hMpc}{h^{-1}\,{\rm Mpc}}

\newcommand{\ie}{i.e.\xspace}
\newcommand{\eg}{e.g.\xspace}

\newcommand{\Lya}{Ly$\alpha$\xspace}

\newcommand{\fesc}{f_{\rm esc}}

\newcommand{\arepo}{{\sc arepo}\xspace}
\newcommand{\areport}{{\sc arepo-rt}\xspace}

\newcommand{\thesan}       {\textsc{thesan}\xspace}
\newcommand{\thesanone}    {\textsc{thesan-1}\xspace}
\newcommand{\thesantwo}    {\textsc{thesan-2}\xspace}
\newcommand{\thesanwc}     {\textsc{thesan-wc-2}\xspace}
\newcommand{\thesanlow}    {\textsc{thesan-low-2}\xspace}
\newcommand{\thesanhigh}   {\textsc{thesan-high-2}\xspace}

\newcommand{\thesansdao}   {\textsc{thesan-sdao-2}\xspace}
\newcommand{\thesantng}    {\textsc{thesan-tng-2}\xspace}

\newcommand{\thesannort}   {\textsc{thesan-nort-2}\xspace}
\newcommand{\thesandarkone}{\textsc{thesan-dark-1}\xspace}
\newcommand{\thesandarktwo}{\textsc{thesan-dark-2}\xspace}
\newcommand{\thesanhr}     {\textsc{thesan-hr}\xspace}
\newcommand{\thesanvhr}    {\textsc{thesan-hr-res8x}\xspace}
\newcommand{\thesanhrlarge}{\textsc{thesan-hr-large}\xspace}
\newcommand{\thesanhrwdm}  {\textsc{thesan-hr-wdm}\xspace}
\newcommand{\thesanhrfdm}  {\textsc{thesan-hr-fdm}\xspace}
\newcommand{\thesanhrsdao} {\textsc{thesan-hr-sdao}\xspace}

\newcommand{\paperI}{Paper~\textsc{I}\xspace}
\newcommand{\paperII}{Paper~\textsc{II}\xspace}
\newcommand{\paperIII}{Paper~\textsc{III}\xspace}

\newcommand{\highz}{\mbox{high-$z$}\xspace}
\newcommand{\disperse}{\textsc{DisPerSE}\xspace}
\newcommand{\filespecs}[3]{\textit{File name}: \texttt{#1}\newline\noindent\textit{Availability}: #3\\}
\newcommand{\dataused}[1]{\textit{Data product used}: #1\\}

\newcommand{\rev}[2]{#2}

\graphicspath{{./}{figures/}}

\title[\thesan data release \& JWST comparison]{The \thesan\ project: public data release of radiation-hydrodynamic simulations matching reionization-era JWST observations}

\author[E.~Garaldi et al.]{%
Enrico~Garaldi\orcidlink{0000-0002-6021-7020}$^{1,2}$\thanks{E-mail: \href{mailto:egaraldi@sissa.it}{egaraldi@sissa.it} (EG); \href{mailto:kannanr@yorku.ca}{kannanr@yorku.ca} (RK); \href{mailto:asmith@utdallas.edu}{asmith@utdallas.edu} (AS); \href{mailto:josh@joshborrow.com}{josh@joshborrow.com} (JB)}, 
Rahul~Kannan\orcidlink{0000-0001-6092-2187}$^{3}$, 
Aaron~Smith\orcidlink{0000-0002-2838-9033}$^{4}$, 
Josh~Borrow\orcidlink{0000-0002-1327-1921}$^{5}$, 
Mark~Vogelsberger\orcidlink{0000-0001-8593-7692}$^{5,6}$,
\newauthor
R\"udiger~Pakmor\orcidlink{0000-0003-3308-2420}$^{1}$,
Volker~Springel\orcidlink{0000-0001-5976-4599}$^{1}$,
Lars~Hernquist$^{7}$,
Daniela~Gal\'arraga-Espinosa\orcidlink{0000-0002-8808-803X}$^{1}$,
\newauthor
Jessica~Y.-C.~Yeh\orcidlink{0000-0002-5721-7679}$^{8}$,
Xuejian~Shen\orcidlink{0000-0002-6196-823X}$^{5,9}$,
Clara~Xu\orcidlink{0000-0003-4688-6487}$^{5}$,
Meredith~Neyer\orcidlink{0000-0002-9205-9717}$^{5}$,
Benedetta~Spina\orcidlink{0000-0003-1634-1283}$^{10}$,
\newauthor
Mouza~Almualla\orcidlink{0000-0002-4694-7123}$^{7}$ 
and
Yu~Zhao\orcidlink{0000-0001-6768-9947}$^{11}$ 
\\%
\\%
$^{1}$Max-Planck Institute for Astrophysics, Karl-Schwarzschild-Str.~1, D-85741 Garching, Germany\\%
$^{2}$ Institute for Fundamental Physics of the Universe, via Beirut 2, 34151 Trieste, Italy\\
$^{3}$Department of Physics and Astronomy, York University, 4700 Keele Street, Toronto, ON M3J 1P3, Canada\\%
$^{4}$Department of Physics, The University of Texas at Dallas, Richardson, Texas 75080, USA\\%
$^{5}$Department of Physics $\&$ Kavli Institute for Astrophysics and Space Research, Massachusetts Institute of Technology, Cambridge, MA 02139, USA\\%
$^{6}$The NSF AI Institute for Artificial Intelligence and Fundamental Interactions, Massachusetts Institute of Technology, Cambridge, MA 02139, USA\\%
$^{7}$Center for Astrophysics $\vert$ Harvard $\&$ Smithsonian, 60 Garden Street, Cambridge, MA 02138, USA\\%
$^{8}$Department of Physics, Stanford University, 382 Via Pueblo Mall, Stanford, CA 94305, USA\\%
$^{9}$TAPIR, California Institute of Technology, Pasadena, CA 91125, USA\\%
$^{10}$Argelander Institut f\"ur Astronomie, Auf dem H\"ugel 71, 53121 Bonn, Germany\\%
$^{11}$Department of Physics and Astronomy, University of Southern California, Los Angeles,
CA 90007, USA%
}

\date{Accepted XXX. Received YYY; in original form ZZZ}

\pubyear{2023}

\begin{document}
\label{firstpage}
\pagerange{\pageref{firstpage}--\pageref{lastpage}}
\maketitle

\begin{abstract}
Cosmological simulations serve as invaluable tools for understanding the Universe. However, the technical complexity and substantial computational resources required to generate such simulations often limit their accessibility within the broader research community. Notable exceptions exist, but most are not suited for simultaneously studying the physics of galaxy formation and cosmic reionization during the first billion years of cosmic history. This is especially relevant now that a fleet of advanced observatories (\eg \textit{James Webb Space Telescope}, \textit{Nancy Grace Roman Space Telescope}, SPHEREx, ELT, SKA) will soon provide an holistic picture of this defining epoch. To bridge this gap, we publicly release all simulation outputs and post-processing products generated within the \thesan simulation project at \href{https://www.thesan-project.com}{www.thesan-project.com}. This project focuses on the $z \geq 5.5$ Universe, combining a radiation-hydrodynamics solver (\areport), a well-tested galaxy formation model (IllustrisTNG) and cosmic dust physics to provide a comprehensive view of the Epoch of Reionization. The \thesan suite includes 16 distinct simulations, each varying in volume, resolution, and underlying physical models. This paper outlines the unique features of these new simulations, the production and detailed format of the wide range of derived data products, and the process for data retrieval. Finally, as a case study, we compare our simulation data with a number of recent observations from the \textit{James Webb Space Telescope}, affirming the accuracy and applicability of \thesan. The examples also serve as prototypes for how to utilise the released dataset to perform comparisons between predictions and observations.
\end{abstract}


\section{Introduction}
\label{sec:intro}

Numerical simulations are a fundamental pillar of modern astrophysical research. They represent the most practical approach to solve the equations governing the complex multi-scale (astro)physics of galaxy formation in a general and coupled way. Moreover, these simulations enable synthetic experiments that sidestep the inherent limitations of conducting real-world tests on astronomical entities, thereby fulfilling a key aspect of the scientific method within astrophysics.

Recent decades have witnessed the rapid transition from pioneering works employing relatively small-scale, low-resolution dark matter (DM)-only simulations \citep{PressSchechter1974, Davis+1985} to significantly more advanced and rigorously validated, large-scale cosmological hydrodynamical simulations \citep[for a recent review of the field see][]{galform_review}. Notable among the latter are: Illustris \citep{Illustris_nature, Illustris, Genel2014}, Eagle \citep{eagle}, MassiveBlackII \citep{massive-black-ii}, Magneticum \citep{magneticum}, Romulus-25 \citep{romulus25}, IllustrisTNG \citep{TNG_Marinacci, TNG_Naiman, TNG_Nelson, TNG_Pillepich, TNG_Springel,Pillepich2019,Nelson2019b}, Simba \citep{Simba}, NewHorizon \citep{newhorizon}, FIREBox \citep{firebox} and MillenniumTNG \citep{MTNG, PakmorMTNG, Kannan2023}. For the first time, these simulations have achieved a comprehensive and accurate representation description of the observed properties of galaxies over a representative volume and a large fraction of the Universe's history. 
They self-consistently follow the evolution of DM and gas, and implement models for the formation and evolution of stellar populations and black holes. However, a common limitation among these simulations is the fact that they all forego simulating radiation and instead employ a spatially-uniform time-evolving radiation field. 
While this \rev{is a good approximation in the post-reionization Universe}{approximation holds reasonably well in the post-reionization inter-galactic medium (IGM)} below $z \lesssim 4.5$, it breaks down at higher redshifts as a consequence of the highly spatially inhomogeneous radiation field emitted by localised, short-lived sources and with short mean free paths. \rev{}{Additionally, such approximation is never accurate inside and in the vicinity of galaxies, where the radiation from local sources dominates over the background for several tens of kpc.}

In recent years, a number of numerical simulations have focused on the Cosmic Dawn and the Epoch of Reionization (EoR). The latter was a pivotal period of time when the radiation emitted by the first stars and galaxies progressively ionized the gas in the IGM between them. Some noteworthy examples in this domain are CROC \citep{CROC}, CoDa \citep{CoDa, CoDaII, CoDaIII}, AURORA \citep{Aurora}, SPHINX \citep{SPHINX,SPHINX20}, OBELISK \citep{Obelisk} and Technicolor Dawn \citep{TechnicolorDawn}, each endeavouring to self-consistently incorporate the effect of radiation transport (RT). However, the computational demands imposed by RT often result in limitations, the most common of which are restricted simulation volumes or too coarse of resolutions. In addition, all these runs employ physical models that have been calibrated at the high redshifts investigated (often by matching the observed UV luminosity function) and not as well-studied past $z\lesssim5$.
Furthermore, these simulations typically rely on physical models calibrated for high-redshift phenomena, often validated against observed UV luminosity functions. This raises questions about the model applicability beyond redshifts $z<5$ and, consequently, the generalizability of the insights derived from them. 
For instance, extending the SPHINX galaxy formation model to lower redshift appears to give rise to an excessive number of bulge-dominated galaxies \citep{Mitchell2021}.

To address some of the limitations outlined above, we have developed the \thesan suite of radiation-magneto-hydrodynamical cosmological simulations, designed to 
provide a comprehensive view of the $z\gtrsim 5$ reionizing Universe \citep[presented in the three introductory papers:][\paperI, \paperII and \paperIII, hereafter]{Thesan_intro, Thesan_igm, Thesan_Lya}. 
\thesan leverages the well-tested IllustrisTNG galaxy formation model, which has been successful in reproducing the properties of post-reionization galaxies, and complements this with self-consistent radiation transport and cosmic dust modelling. The suite follows a volume large enough to capture the average properties of reionization\footnote{\rev{}{The volume required for convergence of reionization properties remains unclear. We discuss this and substantiate our claim in Sec.~\ref{sec:physical_considerations}}}, while also maintaining a resolution adequate to accurately simulate global galaxy properties. 
It features runs that explore a variety of physical models for the escape of ionizing photons and the nature of DM. Today, \thesan stands as one of the most advanced tools available to study the reionizing Universe in a holistic manner, offering reliable predictions across a range of physical quantities related to both galaxies and cosmic reionization. 
To make the \thesan suite an asset to the entire community, we are publicly releasing all of the available raw data and post-processing products. This paper serves as an overview of this data release and aims at providing guidance for practical use.

The practice of publicly releasing large datasets in astronomy has a relatively long history now, both for observational data \citep[starting from the SDSS SkyServer,][]{SDSS_SkyServer} and numerical simulations \citep[\eg the release of the Millenium simulation by][]{Millennium, LemsonVirgo2006}. However, \rev{this openness has not been widely adopted for simulations of the EoR, which have largely remained proprietary, effectively limiting}{while there has been some initial progress, in the context of EoR simulations the adoption of open data practices remains rare, partially due to resource constraints. This effectively limits} their broader utilisation and reproducibility. We aim to establish this as standard practice with our data release, which we \rev{loosely model}{implement through the accessible and widely-used Globus\footnote{\url{https://www.globus.org/}} tool (see Sec. \ref{sec:data_retrieval}). We loosely model the release}{} after the IllustrisTNG \citep[][]{TNG_data_release} \rev{data release}{one}. 
We offer an online interface to selectively download raw data and post-processed data products, and we provide extensive online documentation and usage examples, and the possibility to import \thesan data into popular analysis tools.

Recently, the SPHINX collaboration announced a public data release \citep{SPHINX_release}. However, at the time of writing, the available data consists of a catalogue containing galaxy properties and synthetic spectra for a sub-sample of 1400 simulated galaxies. We expand significantly on this with our unrestricted data release, allowing downloads of our full snapshot and post-processed data at all output times, providing significant scope for analyses focused on dynamics within individual galaxies all the way up to the science of the intergalactic medium.

In this paper, we begin by summarising the salient properties of the \thesan simulations in Sec.~\ref{sec:simulations}. We then provide a thorough description of the released data and associated analysis products in Sec.~\ref{sec:data_products}, including instructions to retrieve them. We discuss some considerations for effective usage in Sec.~\ref{sec:usage}. Finally, we showcase the predictive power and physical richness of the \thesan simulations by providing a few example comparisons with recent \textit{James Webb Space Telescope} (\textit{JWST}) observations in Sec.~\ref{sec:jwst_comparison}, followed by our summary and conclusions in Sec.~\ref{sec:conclusions}.

\section{Simulation description}
\label{sec:simulations}

\begin{table*}
	\centering
	\caption{Overview of the \thesan simulation suite. The columns, from left to right, represent the following attributes: the name of the simulation, linear size of the simulation box, (initial) particle number, mass of DM and gas particles, (minimum) softening length for (gas) star and DM particles, minimum physical cell size at $z=5.5$, final redshift, escape fraction of ionizing photons from the birth cloud (if applicable), and a short description of the simulation.}
	\label{table:simulations}
        \addtolength{\tabcolsep}{1.6pt}
	\begin{tabular}{lccccccccc} 
            \hline
            \vspace{-.25cm} \\
		Name & $L_\mathrm{box}$ & $N_\mathrm{particles}$ & $m_\mathrm{DM}$ & $m_\mathrm{gas}$ & $\epsilon$ & $r^\mathrm{min}_\mathrm{cell}$& $z_\mathrm{end}$ & $f_\mathrm{esc}$ & Description\\  
		& [cMpc] & & [$\mathrm{M}_\odot$] & [$\mathrm{M}_\odot$] & [ckpc] & [pc] & & & \vspace{.1cm} \\
		\hline
            \vspace{-.25cm} \\
		\thesanone & $95.5$  & $2 \times 2100^3$ & $3.12 \times 10^6$ & $5.82 \times 10^5$ & $2.2$ & $10$ & $5.5$ & $0.37$ & Fiducial \\
		&&&&&&&&&\\
		\thesantwo & $95.5$  & $2 \times 1050^3$ & $2.49 \times 10^7$ & $4.66 \times 10^6$ & $4.1$ & $35$ & $5.5$ & $0.37$ & Fiducial\\
		\thesanwc & $95.5$  & $2 \times 1050^3$ & $2.49 \times 10^7$ & $4.66 \times 10^6$ & $4.1$ & $33$ & $5.5$ & $0.43$ & Weak convergence of $x_\mathrm{HI} (z)$ \\
		\thesanhigh & $95.5$  & $2 \times 1050^3$ & $2.49 \times 10^7$ & $4.66 \times 10^6$ & $4.1$ & $33$ & $5.5$ & $0.8$ & $f_{\mathrm{esc}} \propto \mathrm{M}_\mathrm{halo} (> 10^{10})$\\
		\thesanlow & $95.5$  & $2 \times 1050^3$ & $2.49 \times 10^7$ & $4.66 \times 10^6$ & $4.1$ & $32$ & $5.5$ & $0.95$ & $f_{\mathrm{esc}} \propto \mathrm{M}_\mathrm{halo} (<10^{10})$\\
		\thesansdao & $95.5$  & $2 \times 1050^3$ & $2.49 \times 10^7$ & $4.66 \times 10^6$ & $4.1$ & $33$ &  $5.5$ & 0.55 & Strong dark acoustic oscillations\\
		&&&&&&&&&\\
		\thesantng & $95.5$  & $2 \times 1050^3$ & $2.49 \times 10^7$ & $4.66 \times 10^6$ & $4.1$ & $30$ &  $5.5$ & - & Original TNG model\\
        
		\thesannort & $95.5$  & $2 \times 1050^3$ & $2.49 \times 10^7$ & $4.66 \times 10^6$ & $4.1$ & $35$ &  $5.5$ & - & No radiation\\
		&&&&&&&&&\\
		\thesandarkone & $95.5$  & $2100^3$ & $3.70 \times 10^6$ & - & $2.2$ & - &  $0.0$ & - & DM only\\
		\thesandarktwo & $95.5$  & $1050^3$ & $2.96 \times 10^7$ & - & $4.1$ & - &  $0.0$ & - & DM only\\
		&&&&&&&&&\\
		\thesanvhr     & $5.9$   & $2 \times 512^3$ & $6.03 \times 10^{4}$ & $1.13 \times 10^{4}$ & $0.425$ & $8$ & $5.0$ & $0.37$ & Fiducial \\
		\thesanhr      & $5.9$   & $2 \times 256^3$ & $4.82 \times 10^{5}$ & $9.04 \times 10^{4}$ & $0.85$  & $32$ & $5.0$ & $0.37$ & Fiducial \\
		\thesanhrlarge & $11.8$  & $2 \times 512^3$ & $4.82 \times 10^{5}$ & $9.04 \times 10^{4}$ & $0.85$  & $15$ & $5.0$ & $0.37$ & Fiducial \\

		&&&&&&&&&\\
		\thesanhrsdao  & $5.9$   & $2 \times 256^3$ & $4.82 \times 10^{5}$ & $9.04 \times 10^{4}$ & $0.85$  & $33$ & $5.0$ & $0.37$ & Strong dark acoustic oscillations \\
		
		\thesanhrwdm   & $5.9$   & $2 \times 256^3$ & $4.82 \times 10^{5}$ & $9.04 \times 10^{4}$ & $0.85$  & $28$ & $5.0$ & $0.37$ & Warm dark matter ($3\,{\rm keV}$) \\ 
		
		\thesanhrfdm   & $5.9$   & $2 \times 256^3$ & $4.82 \times 10^{5}$ & $9.04 \times 10^{4}$ & $0.85$  & $23$ & $5.0$ & $0.37$ & Fuzzy dark matter ($2\times 10^{-21}\,{\rm eV}$) \vspace{.1cm} \\ 
		
		\hline
	\end{tabular}
        \addtolength{\tabcolsep}{-1.6pt}
\end{table*}

\thesan is a state-of-the-art suite of radiation-magneto-hydrodynamical (RMHD) cosmological simulations, recently developed to provide a holistic view of the early Universe. Specifically, it is designed to simultaneously model the formation of galaxies during the first billion years after the Big Bang, the evolving and spatially-inhomogeneous properties of the IGM during Cosmic Dawn and the EoR, and \textit{crucially} their connection.

To achieve this ambitious objective, \thesan combines a large (95.5\,cMpc)$^3$ volume with sufficiently high resolution to model the formation of atomic cooling haloes, the smallest structures significantly contributing to the reionization process. \thesan simulates a wide range of physical processes relevant to the high-redshift Universe. These include stochastic star formation following the Kennicut-Schmidt relation, magnetic fields, mass, energy and metal return from supernovae (type Ia and II) as well as AGB stars, galactic winds, and black holes (including accretion, bi-modal feedback and radiation output). It also features binary stellar evolution as part of the stellar population synthesis model, explicit tracking of nine metal species (namely H, He, C, N, O, Ne, Mg, Si, Fe), and cosmic dust production, evolution and destruction. In addition, the initial conditions for the \thesan runs were generated using the variance-suppression technique of \citet{AnguloPontzen2016} to boost the statistical significance of box-averaged results.

\subsection{Technical details}
During the design phase of \thesan, we made the explicit decision to utilize physical models that have been calibrated at lower redshifts.
\thesan is an extension of the IllustrisTNG model \citep{Weinberger2017, Pillepich2018b, TNG_Marinacci, TNG_Naiman, TNG_Nelson, TNG_Pillepich, TNG_Springel}, specifically tailored to the reionizing Universe.
As such, it inherits the core physical framework of IllustrisTNG, with the following key modifications:
\begin{itemize}
    \item The spatially-uniform UV background is replaced by self-consistent radiation transport \citep[\areport; ][]{ArepoRT}, which follows photons injected from stars and Active Galactic Nuclei (AGN) across three adjacent frequency bins defined by the following photon energy thresholds $[13.6, 24.6, 54.4, \infty)$\,eV.
    \item The equilibrium cooling of IllustrisTNG is replaced by a non-equilibrium thermo-chemistry module that simultaneously solves the equations for heating, cooling, and ionization, ensuring that the rapidly evolving reionization fronts can be more faithfully tracked.
    \item A model for cosmic dust production, evolution, and destruction by \citet{McKinnon+16, McKinnon+17} is incorporated, facilitating self-consistent predictions of the dust content of high-$z$ galaxies.
\end{itemize}

Adopting this approach offers three distinct advantages: (i) It enables the validation of physical models, originally calibrated at $z\sim0$, in the high-redshift regime. (ii) It ensures that the simulations employ a physical model that remains reliable and realistic even \textit{beyond} the end of the simulations, where it would otherwise be unverifiable\footnote{In principle, despite sharing the same galaxy-formation model, some \thesan galaxies might differ from those produced by the original IllustrisTNG model, since we have upgraded from equilibrium to non-equilibrium cooling in \thesan. However, we have confirmed that this change does not significantly impact the properties of simulated galaxies with a halo mass of $M_\mathrm{halo} \gtrsim 10^{9} \, \Msun$ at the latest available redshift ($z=5.5$). Smaller galaxies are impacted by the self-consistent treatment of RT, and therefore are expected to differ irrespective of the cooling treatment.}. (iii) \thesan introduces only a single additional free parameter, the unresolved stellar escape fraction, $\fesc$. We calibrate the latter by broadly matching the simulated reionization history to a so-called `late reionization' model \citep[\eg][]{Kulkarni2019, Keating+2020}, which best fits the latest observations of \Lya effective optical depth at the tail end of reionization \citep[\eg][]{Zhu+2020, Bosman+2021}. The presence of a singular calibration parameter strengthens the comparison between \thesan and observational data (as done \eg in Sec.~\ref{sec:jwst_comparison}). This serves as a robust test of the physical model implemented in the simulations, which was primarily developed to reproduce low-redshift observations.

The simulations are performed using the \arepo code \citep{Arepo,Pakmor2016,Arepo-public}, which solves the magneto-hydrodynamics equations on a mesh, constructed as the Voronoi tessellation of a set of mesh-generating points that follow the gas flow, providing a natural way to enhance resolution in high-density regions. Gravitational forces are calculated using a hybrid TreePM approach, where the long-range part of the gravitational interaction is calculated using a particle-mesh algorithm, while the short-range forces are computed through a hierarchical oct-tree \citep{Barnes&Hut86}. 
Radiation is included through the \areport extension \citep{ArepoRT}, which solves the first two moments of the RT equation, supplemented by the M1 closure relation \citep{Levermore84}, with a manifestly photon-conserving scheme. The number of photons emitted by stars is determined using the BPASS library \citep[][version 2.1]{BPASS2017, BPASS2018}, assuming a \citet{Chabrier2003} initial mass function (IMF) for stars. We assume the radiation emitted by AGN follows the spectral shape of \citet{Lusso+2015} rescaled to match the total bolometric luminosity computed from the simulated black hole accretion rate. Please refer to \paperI for further details on the design and calibration of the simulations.

\subsection{Numerical setup and available simulations}

\thesan is composed of one flagship run, designated as \thesanone, which features the highest resolution and implements our fiducial physical model. It is complemented by a series of additional runs, labelled as \textsc{thesan-...-2}), with 8 times lower mass resolution exploring different choices for the escape of ionizing photons, the DM model employed, and the numerical convergence of the simulations. In particular, the \thesanlow and \thesanhigh runs are used to investigate the effect of a halo mass-dependent escape fraction of ionizing photons. This is achieved by forcing $\fesc = 0$ in all haloes with mass above and below $10^{10} \, \Msun$, respectively. In \thesansdao, cold DM is replaced with the strong Dark Acoustic Oscillations \citep[sDAO,][]{Cyr-Racine+2014} model by replacing the transfer function used in the production of the initial conditions. \thesanwc provides a run identical to \thesantwo except for the value of $\fesc$, which has been slightly adjusted to compensate the lower star-formation rate density with respect to \thesanone and achieve the same reionization history. Concerning numerical variations, \thesantng implements the original IllustisTNG model (\ie without dust, and with self-consistent RT replaced by a spatially-uniform UVBG), and \thesannort completely neglects radiation. Additionally, we provide two DM-only runs (\thesandarkone and \thesandarktwo) at the same resolution of \thesanone and \thesantwo, but evolved all the way to $z=0$.

All simulations within the \thesan suite follow the evolution of a patch of the Universe with linear size $L_\mathrm{box} = 95.5$~cMpc, starting from the same initial conditions (sampled at different resolutions). In \thesanone, the DM and gas mass resolutions are $m_\mathrm{DM} = 3.12 \times 10^6 \, \Msun$ and $m_\mathrm{gas} = 5.82 \times 10^5 \, \Msun$, respectively. This resolution enables us to resolve atomic cooling haloes (the smallest haloes that contribute significantly to reionization) with approximately 50 DM particles. For the \thesantwo series, these mass resolutions are exactly eight times larger. Table~\ref{table:simulations} provides additional numerical details for each simulation, 
including their name, box length ($L_\mathrm{box}$, in cMpc), initial particle number ($N_\mathrm{particles}$), DM and gas mass resolution ($m_\mathrm{DM}$ and $m_\mathrm{gas}$, respectively, in $\mathrm{M}_\odot$), softening length of star and DM particles ($\epsilon$, in ckpc, also corresponding to the minimum softening length for gas particles), minimum physical cell size at $z=5.5$ ($r^\mathrm{min}_\mathrm{cell}$, in pc), final redshift ($z_\mathrm{end}$), subgrid escape fraction of ionizing photons from the birth cloud (if applicable, $f_\mathrm{esc}$), and a short description of the unique feature of the simulation compared to the other runs in the table.

The numbers reported in Table~\ref{table:simulations} are placed into the broader context of the community effort to simulate the reionizing Universe in Fig.~\ref{fig:sim_size}, where we compare the force resolution and the simulated volume of the different simulations in the \thesan project to those of other radiation hydrodynamic simulations designed to model the formation and evolution of structures at high redshift, namely CROC \citep{CROC}, CoDa \citep{CoDa,CoDaII,CoDaIII}, SPHINX \citep{SPHINX,SPHINX20}, OBELISK \citep{Obelisk}, Aurora \citep{Aurora}, TechnicolorDawn \citep{TechnicolorDawn}, Renaissance \citep{Renaissance}, Phoenix \citep{PhoenixSim} \rev{}{and SPICE \citep{SPICE}}. 
For some of them, we have highlighted peculiarities that set them apart from the others in the same group. This emphasizes how the plotted quantities are only two of the many that might characterise a simulation. 
These simulations range from high-resolution small-volume simulations designed to study the formation of first stars, like the Renaissance and Phoenix simulations, to lower-resolution but large-scale simulations that model a large representative volume of the Universe, namely CROC, CoDa, and \thesan. These large volumes are necessary to obtain converged reionization histories and rigorous large-scale IGM predictions \citep{Iliev+2014,GnedinMadau_review}. In contrast, the smaller volume calculations primarily serve as galaxy formation simulations with coupled RT rather than comprehensive reionization simulations. Another advantage of the \thesan simulations is that they use a well-tested galaxy formation model that has been shown to reproduce realistic galaxy properties down to redshift zero \citep{TNG_Springel}. In contrast, some other simulations use galaxy formation models that have been designed primarily for higher redshifts. When these galaxy formation models are used to simulate galaxies down to lower redshifts, they give rise to galaxy populations that are incompatible with observations. An example is the model used in SPHNIX and Obelisk, which produces galaxies with much higher stellar masses than observed, that are bugle-dominated, and have centrally peaked rotation curves \citep{Mitchell2021}. These drawbacks make it difficult to assess the accuracy of certain predictions like the LyC production rate and emission line properties, which might be overestimated due to the high stellar masses, and the escape fraction of ionizing photons, which might be underestimated due to inefficient stellar feedback \citep{Trebitsch+2020}. Therefore, the \thesan simulations offer a distinctively comprehensive and reliable approach to the simultaneous study of structure formation and reionization at high redshift.

\subsection{Thesan-HR}
\begin{figure}
    \centering
    \includegraphics[width=\columnwidth]{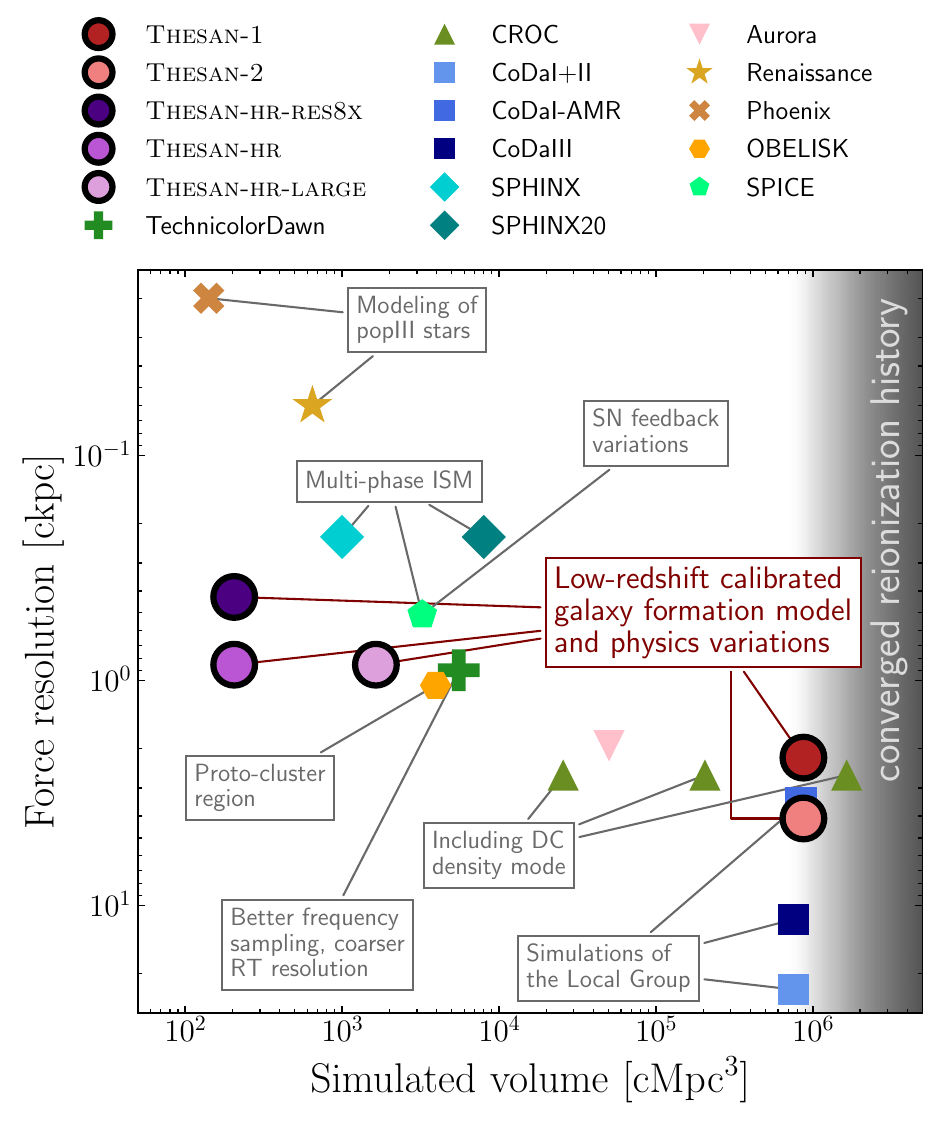}
    \caption{Overview of simulated volume and force resolution of R(M)HD simulations. For some of them we have highlighted peculiarities that set them apart from otherwise similar runs. For grid-based codes, when not otherwise indicated by the authors, we consider the force resolution to be equal to 3 times the smallest grid cell size. The background grey shading indicates simulation volumes large enough to produce a converged reionization history, making them reliable for reionization studies. The colour gradient reflects the range of volumes required according to different studies \citep{Iliev+2014, GnedinMadau_review}.}
    \label{fig:sim_size}
\end{figure}

The \thesan suite has been recently augmented with a specialized set of higher-resolution, small-box simulations, designated as \thesanhr. These runs are designed to study in detail the impact of inhomogeneous radiation propagation on smaller haloes. This is carried out both within the framework of the standard \LCDM cosmology \citep{ThesanHR} and in beyond-CDM models \citep{ThesanHR_noncdm}. We release the data for these runs alongside the main \thesan boxes, and include them in both Table~\ref{table:simulations} and Fig.~\ref{fig:sim_size} for catalog and visual representations.

\subsection{Overview of the first results}
\label{sec:results_overview}
The \thesan suite has already produced a number of scientific results. We summarise here those that could potentially impact future applications of the data. Specifically, we have calibrated the birth cloud escape fraction of ionizing photons ($\fesc$) to align with a late reionization history in \thesanone. This calibrated value is consistently applied across \thesantwo and \thesansdao. In \thesanwc, the value has been fine-tuned to match the reionization history of \thesanone. For \thesanlow and \thesanhigh, it has been re-calibrated due to their modified mechanisms for the escape of ionizing photons. This procedure results in all of the \thesan runs reaching a \rev{}{volume-weighted}\footnote{\rev{}{If not otherwise specified, all ionised fraction quoted in the paper are volume weighted, as customary in studies of cosmic reionization.}} hydrogen ionized fraction of 99\% by $z \lesssim 5.8$, with \thesanone reaching this value at $z \approx 5.5$ (see Fig.~3 of \paperII). The only exception is \thesanlow, which completes reionization much earlier (approximately at $z \approx 6.6$) as a result of the enhanced contribution from low-mass galaxies that dominate at higher redshifts.

\thesan exhibits strong agreement with observed stellar-to-halo-mass relations and stellar mass functions at $6 \leq z \leq 10$ (see Fig.~9 and 10 of \paperI). It also aligns well with the high-redshift UV luminosity function over the same redshift range (UVLF, Fig.~11 of \paperI). This is remarkable, as this quantity is typically calibrated against in reionization-era simulations. It should be noted, however, that to reach such an agreement it is necessary to adopt the dust attenuation relation developed for IllustrisTNG in \citet{Vogelsberger2020}. Using the self-consistently simulated dust masses, the bright end of the UVLF is not attenuated enough, suggesting a potential under-production of dust in \thesan. Unfortunately, observations in the same stellar mass range only provide upper limits on the total dust mass (see Fig.~15 of \paperI). The metal production, however, is in excellent agreement with the observed mass-metallicity relation (see Fig.~14 of \paperI).

The IGM in \thesan shows a realistic broad distribution in the temperature-density space, both during and after reionization (see Fig.~4 of \paperII). Key metrics such as the HI photo-ionization rate in ionized regions, the mean free path of ionizing photons, and the CMB optical depth are in excellent agreement with existing data (see Fig.~5, 7 and 3 of \paperII, respectively). However, the \Lya mean transmitted flux and effective optical depths distribution consistently indicate that a slightly later reionization is preferred (see Fig.~9 and 11 of \paperII). \rev{}{While Fig.~2 of \paperII might appear to indicate the opposite, we caution the reader that the observational values of $x_\mathrm{HI}$ at the tail-end of reionization are typically derived by mapping the observed \Lya flux to HI fractions through simulations employing cruder approximations than \thesan. Therefore, we consider a comparison with the \Lya flux a more direct test of our model, and trust the conclusion drawn from it more than those from a comparison of the simulated and observationally-inferred $x_\mathrm{HI}$.} Finally, \thesan is the first simulation of its kind to reproduce the excess transmitted flux in proximity of galaxies in the \Lya forest of high-redshift quasar spectra (see Fig.~15 of \paperII).

The \thesan suite has also been used to study the production of \Lya photons and their transmission through the CGM of galaxies (\paperIII), to provide predictions for forthcoming intensity mapping experiments \citep[][]{Thesan_lim}, to study the escape of ionizing photons from galaxies \citep{Thesan_fesc}, to calibrate effective field theory models and predict the 21cm signal of neutral hydrogen \citep{Thesan_eft}, to study the  \Lya emitter luminosity function \citep{Thesan_LAE_LF}, to deepen the understanding of how a UV background influences galaxy properties \citep{ThesanHR}, to predict high-redshift observables in beyond-CDM models \citep{ThesanHR_noncdm}, \rev{}{to compare the evolution of the ionised bubble size to available observations \citep{Thesan_bubbles}, to investigate the evolution of galaxy sizes \citep{Thesan_sizes}} and --~combined with the IllustrisTNG and MilleniumTNG runs~-- to provide predictions of the $z\gtrsim6$ galaxy properties over a wide range of scales \citep{Kannan2023}. All these studies are based on the data that we are publicly releasing alongside this paper, and will we describe in detail in the following sections.

\begin{table*}
    \centering
    
    \caption{Overview of the available data and data products for each of the \thesan simulations. For each simulation indicated in the first column, a \cmark\xspace indicates that the data product referenced in the corresponding column is available, while a \xmark\xspace signifies that these data are not available. The data products are thoroughly described in Sec.~\ref{sec:data_products}. In the future, more data will be made publicly available.}
    \label{table:status}
    \begin{tabular}{l|cccccccccccc}
    \hline
        Name           & Snapshots & Cartesian  & Merger & Offsets & Hashtables &  SED & Filaments &           \multicolumn{5}{c}{catalogs}                \\ \cline{9-13}
                       &           & outputs     & trees  &         &            &      &           & Groups & \Lya   & cross-link & LOS         & $\fesc$ \\
        \hline
        \thesanone     & \cmark    & \cmark    & \cmark   & \cmark  & \cmark   & \cmark & \cmark    & \cmark & \cmark & \cmark      & \cmark      & \cmark  \\
        \\     
        \thesantwo     & \cmark    & \cmark    & \cmark   & \cmark  & \cmark   & \xmark & \cmark    & \cmark & \cmark & \cmark      & \cmark      & \cmark  \\
        \thesanwc      & \cmark    & \cmark    & \cmark   & \cmark  & \cmark   & \xmark & \cmark    & \cmark & \cmark & \cmark      & \cmark      & \cmark  \\
        \thesanhigh    & \cmark    & \cmark    & \cmark   & \cmark  & \cmark   & \xmark & \cmark    & \cmark & \cmark & \cmark      & \cmark      & \cmark  \\
        \thesanlow     & \cmark    & \cmark    & \cmark   & \cmark  & \cmark   & \xmark & \cmark    & \cmark & \cmark & \cmark      & \cmark      & \cmark  \\
        \thesansdao    & \cmark    & \cmark    & \cmark   & \cmark  & \cmark   & \xmark & \cmark    & \cmark & \cmark & \cmark      & \cmark      & \cmark  \\
        &&&&&&&&&&&\\
        \thesantng     & \cmark    & \xmark    & \cmark   & \cmark  & \cmark   & \xmark & \cmark    & \cmark & \xmark & \cmark      & \cmark      & \xmark  \\
        \thesannort    & \cmark    & \xmark    & \cmark   & \cmark  & \cmark   & \xmark & \cmark    & \cmark & \xmark & \cmark      & \cmark      & \xmark  \\
        &&&&&&&&&&&\\
        \thesandarkone & \cmark    & \xmark    & \cmark   & \cmark  & \xmark   & \xmark & \xmark    & \cmark & \xmark & \cmark      & \xmark      & \xmark  \\
        \thesandarktwo & \cmark    & \xmark    & \cmark   & \cmark  & \cmark   & \xmark & \xmark    & \cmark & \xmark & \cmark      & \xmark      & \xmark  \\
        &&&&&&&&&&&\\
        \thesanvhr     & \cmark    & \xmark    & \xmark   & \cmark  & \xmark   & \xmark & \xmark    & \cmark & \cmark & \xmark      & \xmark      & \cmark  \\
        \thesanhr      & \cmark    & \xmark    & \cmark   & \cmark  & \xmark   & \xmark & \xmark    & \cmark & \cmark & \xmark      & \xmark      & \cmark  \\
        \thesanhrlarge & \cmark    & \xmark    & \cmark   & \cmark  & \xmark   & \xmark & \xmark    & \cmark & \cmark & \xmark      & \xmark      & \cmark  \\
        \thesanhrsdao  & \cmark    & \xmark    & \xmark   & \cmark  & \xmark   & \xmark & \xmark    & \cmark & \cmark & \xmark      & \xmark      & \cmark  \\
        \thesanhrwdm   & \cmark    & \xmark    & \xmark   & \cmark  & \xmark   & \xmark & \xmark    & \cmark & \cmark & \xmark      & \xmark      & \cmark  \\
        \thesanhrfdm   & \cmark    & \xmark    & \xmark   & \cmark  & \xmark   & \xmark & \xmark    & \cmark & \cmark & \xmark      & \xmark      & \cmark  \\
        \hline
    \end{tabular}
\end{table*}

\section{Data and data products}
\label{sec:data_products}

\begin{figure*}
    \centering
    \includegraphics[width=0.92\textwidth]{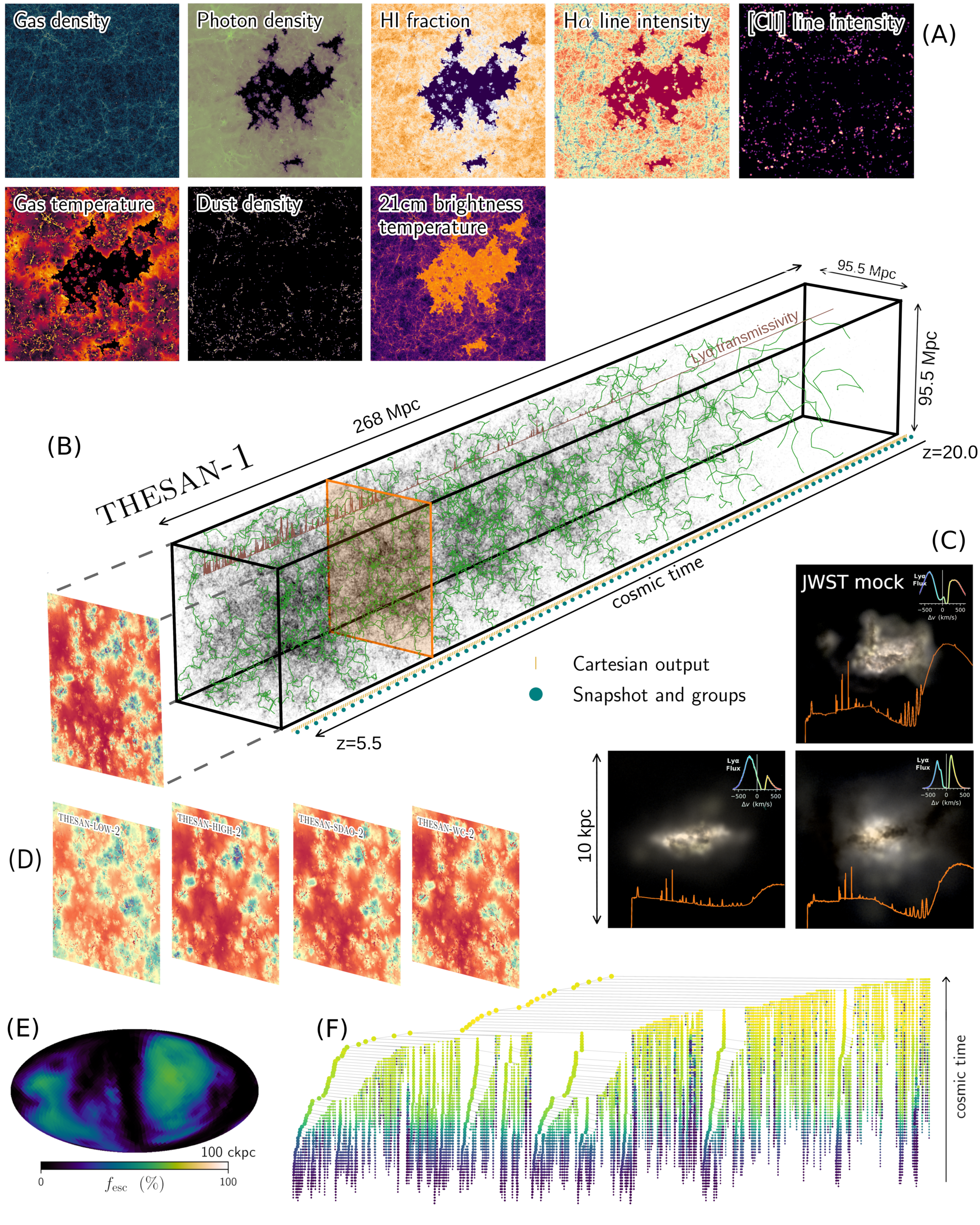}
    
    \caption{Schematic overview of select data products released with this paper and showing their interconnection. Panel (B) features a lightcone-like 3D rendering of the \thesanone simulation box in the redshift range $5.5 \leq z \leq 20$. The simulation box is outlined by black lines, and the output times for full snapshots (see Sec.~\ref{sec:snapshots}) and Cartesian outputs (see Sec.~\ref{sec:cartesian}) are indicated in the bottom right of the panel using blue points and yellow bars, respectively. Gray points within the 3D rendering represent simulated haloes (see Sec.~\ref{sec:fof_catalogs}). We overplot in green the identified filaments (see Sec.~\ref{sec:filaments_catalogs}), in brown the IGM transmissivity to the \Lya line along a random sightline (see Sec.~\ref{sec:los}), and in orange a slice through the simulation. In panel (A) we show a selection of simulated physical quantities, indicated in the top left of each slice. In Panel (C) we show synthetic \textit{JWST} images and SEDs (background and orange line, respectively, see Sec.~\ref{sec:sed_catalogs}) of three randomly-selected galaxies at $z=6$. Panel (D) shows the reionization redshift computed for the different runs of the original \thesan series. Finally, panels (E) and (F) show, respectively, the escape fraction of ionized photons (see Sec.~\ref{sec:fesc_catalogs}) at 100\,ckpc from the central galaxy and the merger tree (see Sec.~\ref{sec:merger_trees}) for the largest halo in the \thesanone box. In the latter, each progenitor is shown using a circle of size proportional to its mass and colour proportional to the reionization state of its environment (defined as a 1\,cMpc sphere around it).}
    \label{fig:overview}
\end{figure*}

With this paper, we release all the raw data and many post-processing products generated within the \thesan suite until now, amounting to nearly 400~terabytes. We plan to release additional data products in the future, as they become available. Current and future data can be found online at \url{https://thesan-project.com}, alongside with their complete documentation. 
At the time of this first data release, we offer the following data and derived products:

\begin{itemize}
    \item \textbf{Snapshots               } (Sec.~\ref{sec:snapshots})         : Full simulation output at 80 different cosmic times, between $z=20$ and $z=5.5$ (produced approximately every 11 Myr).
    \item \textbf{Group catalogs          } (Sec.~\ref{sec:fof_catalogs})      : Properties of haloes and galaxies in the simulation, at the same 80 cosmic times of the snapshots.
    \item \textbf{Cartesian outputs       } (Sec.~\ref{sec:cartesian})         : Gas properties sampled on a Cartesian grid at high time cadence (approximately every $2.8$ Myr).
    \item \textbf{Simulation.hdf5 file    } (Sec.~\ref{sec:simulationhdf5})    : Utility file linking together chunked datasets of the simulation;
    \item \textbf{Merger trees            } (Sec.~\ref{sec:merger_trees})      : Evolutionary tracks of subhaloes and galaxies in the simulation and their properties.
    \item \textbf{Offset files            } (Sec.~\ref{sec:offsets})           : Utility files to simplify the navigation of snapshot, groups and merger trees.
    \item \textbf{Hashtables              } (Sec.~\ref{sec:hashtables})        : Utility files to simplify the selection and loading of particles based on their spatial position.
    \item \textbf{Cross-link files        } (Sec.~\ref{sec:cross-link})       : Utility files to simplify the identification of `twin' haloes across the different \thesan simulations.
    \item \textbf{Lya catalogs            } (Sec.~\ref{sec:lya_catalogs})      : \Lya properties of galaxies in the simulations.
    \item \textbf{Lines of sight          } (Sec.~\ref{sec:los})               : Collection of gas properties along random lines of sight through the simulation.
    \item \textbf{Filament catalogs       } (Sec.~\ref{sec:filaments_catalogs}): Filaments identified in the simulation and their properties.
    \item \textbf{Escape fraction catalogs} (Sec.~\ref{sec:fesc_catalogs})     : Escape fraction properties of galaxies in the simulation.
    \item \textbf{Galaxy SED catalogs     } (Sec.~\ref{sec:sed_catalogs})      : Synthetic SEDs for well-resolved, star-forming galaxies.
\end{itemize}

In the subsequent subsections, we briefly describe each data product. For each, we specify two key attributes: a template file name 
and a list of runs for which the data are available. It should be noted that not all data products are applicable to every \thesan run. Table~\ref{table:status} shows which data are available for each individual run. In the remainder of this Section, we use the following naming convention: \texttt{NNN} represents the snapshot number, zero-padded on the left to have exactly three digits, while \texttt{C} denotes the unpadded chunk number.

In Fig.~\ref{fig:overview}, we provide an overview of the breadth of the data released. In the central panel B we feature a 3D visualization of simulated haloes (grey points, with transparency proportional to their total mass) within the redshift range $5.5 \leq z \leq 20$ in the \thesanone simulation box (outlined by black lines). The output times for snapshots and Cartesian outputs are marked along the redshift axis using blue points and yellow bars, respectively. Additionally, cosmic filaments identified in the simulation are depicted with green lines, while the simulated IGM transmissivity to the \Lya line along a random sightline is represented in brown, and a cross-sectional slice through the simulation in orange. In panel A we show a number of quantities computed along this slice using Cartesian outputs and synthetic SEDs. Examples of the latter are also shown in Panel C, along with synthetic \textit{JWST} images of the corresponding simulated galaxies (generated using the \texttt{skirt} code; see Sec.~\ref{sec:sed_catalogs}) and the profile of the escaping \Lya line, where the null velocity offset corresponds to the intrinsic galaxy redshift. Panel D presents the reionization redshift computed for different runs of the original \thesan series along a slice through the simulation volume. Lastly, panels (E) and (F) illustrate, respectively, the escape fraction of ionizing photons at 100\,ckpc from the central galaxy and the merger tree for the largest halo in the \thesanone box. In the merger tree, each progenitor is represented by a circle, the size of which is proportional to its mass, and the colour indicates the reionization state of its environment, defined as a 1\,cMpc sphere around it.

\textit{Comparison with IllustrisTNG data}. The data structure for snapshots, group catalogs, merger trees, and offset files are mostly identical in format to those of the IllustrisTNG simulation, allowing the tools developed for these simulations to be used for \thesan as well. However, the additional physics simulated in \thesan necessitates the inclusion of further physical quantities, mainly related to radiation and cosmic dust. We also introduce an additional output type, named Cartesian output (see Sec.~\ref{sec:cartesian}). Conversely, some physical quantities have been omitted from the output, as they are not pertinent to the redshift range covered by \thesan. We highlight new or modified data fields in the accompanying documentation.

\subsection{Snapshots}
\label{sec:snapshots}
\filespecs{snapshot\_NNN/snap\_NNN.C.hdf5}{output/}{all runs}

\begin{table}
    \caption{Correspondence between particle types and the physical components they simulate. Particle types 2 and 3 are not used in \thesan.}
    \label{tab:ptype}
    \centering
    \begin{tabular}{c|l}
    \hline
    \texttt{PartType} & Component \\
    \hline
    0 & Gas \\
    1 & Dark matter \\
    4 & Stars and wind particles \\
    5 & Black holes \\
    \hline
    \end{tabular}
\end{table}

Snapshots contain the most complete description available of the simulation. They host a large number of properties for each resolution element (gas, DM, stars, and black holes) in the simulation. Each snapshot provides a record of the simulation at a given cosmic time. \thesan provides 80 of these snapshots, approximately every 11 Myr, between $z=20$ and $z=5.5$ (the exact output times can be found in the \thesan website). 

\begin{figure}
    \centering
    
    \includegraphics[width=\columnwidth]{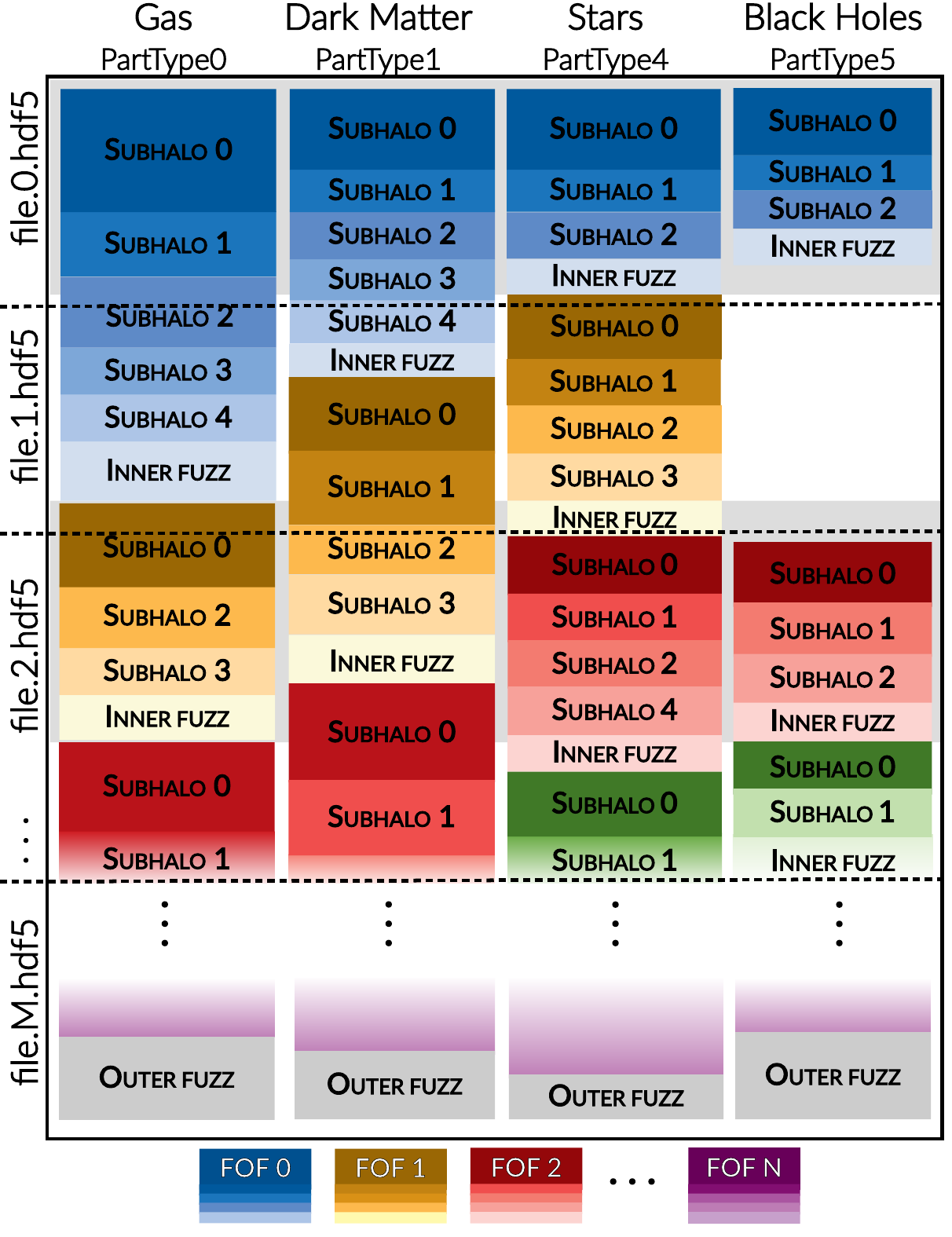}
    \caption{Schematic example of the organization of the \thesan snapshot files. Different colours represent each of the different $N$ FOF groups identified in the simulation. Each row shows a different one of the $M$ total file chunks, with each column representing one of the four different particle types.}
    \label{fig:snapshot_organization}
\end{figure}

Snapshots are stored using the HDF5 file format. 
Because of their size, each snapshot is split into a number of \textit{chunk} files, identical in structure, containing only a portion of the complete dataset. The organisation of a single snapshot is shown in Fig.~\ref{fig:snapshot_organization} and described in the following. Within the snapshot files, particles are divided by type, each one representing different physical components as listed in Table~\ref{tab:ptype} and saved in a different HDF5 group named \texttt{PartTypeP}, where \texttt{P} is an integer. \textit{Within each particle group}, particles are sorted according to the FoF group (see Sec.~\ref{sec:fof_catalogs}) they belong to. This means that all the particles belonging to the FoF group 0 are first, followed by those part of the FoF group 1, and so on. Particles not belonging to any FoF group are last, in the so-called `outer fuzz'. Within each FoF group, particles are sorted according to their subhalo (as identified by SUBFIND). Particles not associated to any subhalo are last within the group (the so-called `inner fuzz'). Finally, within each subhalo, particles are sorted according to their binding energy, from the most bound to the least bound\footnote{This organization allows one to easily locate the particles belonging to a given FOF group or subhalo from the knowledge of just the \textit{number} of particles belonging to each subhalo in the simulation, optimizing the storage by circumventing the need to save the particles identifiers. In order to speed up the particle loading, we have created auxiliary files (namely, the offset files described in Sec.~\ref{sec:offsets} for a halo- or subhalo-based loading, and the hashtables described in Sec.~\ref{sec:hashtables} for a position-based loading).}.  
Once these particle lists containing all particles in the simulations have been built, they are split at arbitrary points and stored in separate chunk files. The splitting is performed in such a way to have approximately the same number of resolution elements in each chunk file. 
As Fig.~\ref{fig:snapshot_organization} shows, if a subhalo (of a FOF group) does not contain particles of a given type, it will simply be skipped. Additionally, there is no guarantee that particles of the same subhalo (of a FOF group) will be stored in the same chunk, or that all the particles of a given type of a subhalo will be stored in the same chunk. Finally, chunks are not guaranteed to contain all particle types. If a particle type is absent in a chunk (as \texttt{PartType5} in chunk 1 in the Figure), the corresponding HDF5 group will not be present in the file. 

Within each \texttt{PartTypeP} HDF5-group, a number of HDF5 datasets are stored, each one corresponding to a different quantity (\eg position, mass, star-formation rate) associated with the resolution elements. The ordering of the particle/cells is the same across these datasets, in such a way that the $i$-th item of each dataset always corresponds to the $i$-th resolution element. A list and description of available datasets are provided 
on the \thesan website. Each dataset contains some attributes that describe the physical quantity stored in the dataset and how it scales with the units of the simulation. 
Finally, each chunk file contains a header group, whose attributes contain general information about the simulation setup, the current snapshot and the current chunk file (also documented on the \thesan website). 

Note that the snapshots contain a number of new quantities not present in IllustrisTNG, that we list and briefly describe in Table~\ref{tab:new_snap_fields}, while some other fields have been dropped, as they are either irrelevant for the redshift covered by \thesan or too memory-consuming. Of particular relevance for the science case of \thesan are the ionized fractions of hydrogen and helium, denoted as \texttt{HI\_Fraction}, \texttt{HeI\_Fraction}, and \texttt{HeII\_Fraction}). These are calculated through the non-equilibrium thermochemistry solver fully coupled with the radiation transport scheme, as opposed to relying on an approximate spatially-uniform UV background. Similarly, \thesan provides the photon number density across the three frequency bins tracked (see Sec.~\ref{sec:simulations}), which is stored in the \texttt{PhotonDensity} dataset in units of $10^{63} \, (\mathrm{ckpc}\,h)^{-3}$. Lastly, the \texttt{GFM\_DustMetallicity} dataset contains the dust-to-gas mass ratio given by the on-the-fly dust model.

\begin{table*}
    \caption{New and modified datasets (with respect to IllustrisTNG) stored in the \thesan snapshots, sorted by particle type.}
    \label{tab:new_snap_fields}
    \centering
    \begin{tabularx}{0.99\textwidth}{c|l|c|p{0.72\linewidth}}
    \hline
    &\texttt{Dataset} & Units & Description \\
    \hline
    \parbox[t]{2mm}{\multirow{7}{*}{\rotatebox[origin=c]{90}{\texttt{PartType0}}}}
    &\texttt{GFM\_DustMetallicity} & - & \emph{New}. Dust-to-gas ratio of gas cell. \\
    &\texttt{GFM\_Metallicity} & - & \emph{Modified}. It now includes metals locked in dust. \\
    &\texttt{GFM\_Metals} & - & \emph{Modified}. It now includes metals locked in dust. Untracked metals are not stored. \\
    &\texttt{HI\_Fraction} & - & \emph{New}. HI fraction of the cells. \\
    &\texttt{HeI\_Fraction} & - & \emph{New}. HeI fraction of the cells. \\
    &\texttt{HeII\_Fraction} & - & \emph{New}. HeII fraction of the cells. \\
    &\texttt{PhotonDensity} & $10^{-63}$ $h^3$ kpc$^{-3}$ & \emph{New}. Density of photons in the gas cell in each spectral band. \\
    \hline
    \parbox[t]{2mm}{\multirow{4}{*}{\rotatebox[origin=c]{90}{\texttt{PartType4}}}}
    &\texttt{BirthDensity} & $10^{10}$ M$_\odot$ $h^2$ ckpc$^{-3}$ & \emph{New}. Comoving mass density where this star particle initially formed. \\
    &\texttt{GFM\_DustMetallicity} & - & \emph{New}. Dust-to-gas ratio of the stellar particle, inherited from the parent gas cell and never changed. \\
    &\texttt{GFM\_Metallicity} & - & \emph{Modified}. It now includes metals locked in dust. \\
    &\texttt{GFM\_Metals} & - & \emph{Modified}. It now includes metals locked in dust. Untracked metals are not stored. \\
    \hline
    \parbox[b]{2mm}{\multirow{3}{*}{\rotatebox[origin=c]{90}{\texttt{PartType5}$\,\,\,\,$}}}
    &\texttt{BH\_MPB\_CumEgyHigh} & \makecell{$10^{10}$ M$_\odot$ $h^{-1}$ \\[-2pt] ckpc$^{2}$ / (0.978 Gyr)} & \makecell[l]{\emph{New}. Cumulative amount of kinetic AGN feedback energy injected into surrounding gas in the high \\[-2pt] accretion-state (quasar) mode along the main progenitor branch of the black hole merger tree.}\\
    &\texttt{BH\_MPB\_CumEgyLow} & \makecell{$10^{10}$ M$_\odot$ $h^{-1}$ \\[-2pt] ckpc$^{2}$ / (0.978 Gyr)} & \makecell[l]{\emph{New}. Cumulative amount of kinetic AGN feedback energy injected into surrounding gas in the low \\[-2pt] accretion-state (wind) mode along the main progenitor branch of the black hole merger tree.}\\
    &\texttt{BH\_PhotonHsml} & $h^{-1}$ ckpc & \emph{New}. Comoving radius enclosing $64 \pm 4$ nearest gas cells, used for photon injection. \\
    \hline
    \end{tabularx}
\end{table*}

\subsection{Halo and subhalo catalogs}
\label{sec:fof_catalogs}
\filespecs{groups\_NNN/fof\_subhalo\_tab\_NNN.C.hdf5}{output/}{all runs}

Whenever a snapshot is produced, we also produce halo (\ie FoF groups) and subhalo (\ie SUBFIND subgroups) catalogs, identified on-the-fly using the FOF and SUBFIND algorithms \citep{Springel+2001}, respectively. The former identifies groups based on spatially close DM particles, while the latter determines groups of physically bound particles. Briefly, the FOF algorithm isolates groups (haloes) of DM particles that are closer than 0.2 times the mean inter-particle distance. Non-DM particles are then associated with the group that contains their nearest DM particle. This subdivision is purely geometrical and does not consider any further physical property of the involved particles and cells. Each of these groups is subsequently  split into gravitationally bound systems (subhaloes) using the SUBFIND algorithm. 

The group catalogs contain a list of identified haloes and subhaloes (stored in the same file, but contained in the \texttt{Group} and \texttt{Subhalo} HDF5-groups, respectively), and their associated properties (stored in separate HDF5-datasets within the corresponding HDF5-group). Similarly to snapshots, these catalogs are split across a number of chunks and are stored using the HDF5 data format. The former are ranked by DM mass (from the most massive to the least massive one), while subhaloes are ranked by parent group first, and then by DM mass within each parent group.

\subsection{Cartesian outputs}
\label{sec:cartesian}
\filespecs{cartesian\_NNN/cartesian\_NNN.C.hdf5}{output/}{all RMHD runs except the \thesanhr series}

\thesan provides a new type of output with respect to IllustrisTNG. Specifically, we save selected gas quantities on a regular Cartesian grid with high time cadence, using a first-order Particle-In-Cell binning approach. \thesanone uses a $1024^3$ grid while the \thesantwo runs use $512^3$ grids, setting the cell sizes to $\sim93$ and $\sim186$ ckpc, respectively. Cartesian outputs are not available for the \thesanhr runs. 
The three-dimensional grid is flattened into a C-ordered array (with the $z$-axis increasing fastest) and then split into chunks, each saved in a different HDF5 file. Each chunk file also stores information as HDF5 attributes of its `Header' HDF5 group, with full details given in the online documentation. 
While most of the quantities are deposited on a regular spatial grid, some observables (\eg the 21 cm emission) are binned in redshift space, accounting for Doppler shifting due to the peculiar velocity of individual gas cells. This redshift-space binning is carried out assuming the observed direction is aligned with the +$z$ axis. The Cartesian outputs are written every $\sim2.8$ Myr amounting to about 400 snapshots over the duration of the simulation. These lower-resolution higher-time-cadence outputs are ideal for studying the large-scale properties of the reionization process.

\subsection{The \texttt{simulation.hdf5} file}
\label{sec:simulationhdf5}
\filespecs{simulation.hdf5}{/}{all runs}

Similar to IllustrisTNG, we produce a `summary' file named \texttt{simulation.hdf5} for each simulation, placed in its root directory. This file employs HDF5 virtual datasets to link all halo and subhalo catalog fields, snapshot particle fields, individual snapshot headers and Cartesian outputs in a single file of limited size. In addition, it also contains configuration flags and parameters of the simulation. Because of its nature, this file requires that all files linked (namely group, snapshot, Cartesian and offsets files) to be downloaded and placed in the same directory structure. This file is completely optional and provided only for convenience. 

The \texttt{simulation.hdf5} file hides the division of (some) outputs into chunks. In fact, through this file it is possible to index fields as if they were not split among file chunks. The virtual datasets take care of loading the requested values from the file chunks and joining them together.

\subsection{Merger trees}
\label{sec:merger_trees}
\filespecs{trees\_sf1\_080.C.hdf5}{postprocessing/trees/LHaloTree}{all runs except the \thesanhr non-\LCDM ones}

We provide merger trees describing the build-up of haloes over cosmic time. These are built using the {\textsc{LHaloTree}} code \citep{Springel+2005} and are named \texttt{trees\_sf1\_080.C.hdf5}, where \texttt{C} is the chunk number (without padding). In short, to determine the appropriate descendant, the unique IDs that label DM particles are tracked between outputs. For a given halo, the algorithm finds all haloes in the subsequent output that contain some of its particles. These are then counted in a weighted fashion, giving higher weight to particles that are more tightly bound in the halo under consideration, and the one with the highest count is selected as the descendant. In this way, preference is given to tracking the fate of the inner parts of a structure, which may survive for a long time upon infall into a bigger halo, even though much of the mass in the outer parts can be quickly stripped. To allow for the possibility that haloes may temporarily disappear for one snapshot, the process is repeated for snapshot $n$ to snapshot $n+2$. If either there is a descendant found in snapshot $n+2$ but none found in snapshot $n+1$, or, if the descendant in snapshot $n+1$ has several direct progenitors and the descendant in snapshot $n+2$ has only one, then a link is made that skips the intervening snapshot. 

The merger tree structure is split across a number of HDF5 files and, within each file, in a number of \texttt{TreeX} groups. For each halo \textit{across all snapshots}, there is one entry in this structure. Haloes are linked together in branches, describing which progenitor haloes have merged to create a descendant one. Each entry of the tree structure contains a link to the descendant (\ie the halo it contributed the most particles in the subsequent snapshot), the next progenitor (\ie the next item in the list of progenitors at the same redshift that contributed to the descendant halo) and its first progenitor (\ie the most massive halo in the list of its progenitor in the previous snapshot). They also contain a number of subhalo properties inherited from the group catalog that are stored in these files for ease of access.

\subsection{Offset files}
\label{sec:offsets}
\filespecs{offsets\_NNN.hdf5}{postprocessing/offsets}{all runs}

The ordering of particles in the snapshots as well as the large number of particles and haloes in the simulation render it sometimes inconvenient or slow to load specific particles, haloes or merger tree entries. To ease this task, we produced so-called offset files, designed to allow an easy navigation between other data products. There is one such file for each snapshot of each simulation. These files provide a mapping between halo or particle indices and their memory position (i.e. chunk file and position within it), as well as a mapping between the halo number and position in the merger tree structures (identified as a combination of chunk number, tree number and index within the tree). The offset files provide also the inverse mapping, from chunk file to the range of particles or groups contained.

The \thesan website contains a thorough description of available fields in these files. 
Practically speaking, in order to, \eg, load the gas particles of the subhalo with global index $j$ from the simulation snapshots, one can simply:
\begin{itemize}
    \item load the offset file for the relevant redshift,
    \item index the HDF5 dataset \texttt{Subhalo/SnapByType} with the index $j$ to obtain the global index $k$ of the \textit{first} gas particle of the subhalo,
    \item index the HDF5 dataset \texttt{FileOffsets/SnapByType} with the index $k$ to obtain the chunk number $n$ storing the particle,
    \item load the appropriate chunk file $n$ and load the subhalo gas particles, potentially continuing from the following chunk file ($n+1$) if the chunk file $n$ contains less gas particles than belonging to the subhalo (as indicated by the \texttt{Subhalo/SubhaloLenType} HDF5 dataset in the group catalog).
\end{itemize}

We provide utility \texttt{python} functions to perform this and other tasks on the \thesan website.

\subsection{Hashtables}
\label{sec:hashtables}
\filespecs{spatial\_hashtable\_NNN.hdf5}{postprocessing/hashtables}{all runs except the \thesanhr series}

\begin{figure}
    \centering
    \includegraphics[width=\columnwidth]{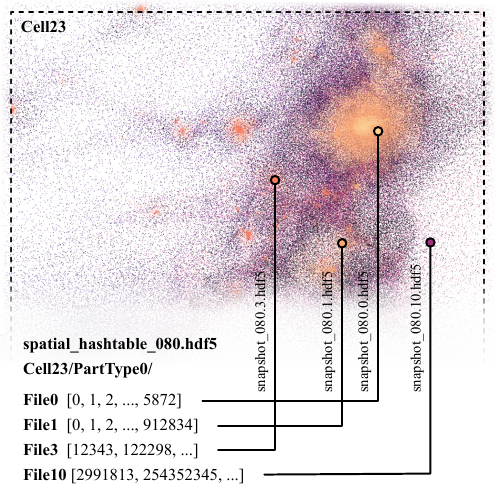}
    \caption{A demonstration of the spatial hashtable applied to \thesan data. Each point represents one gas (particle type 0) particle, with the points coloured by the sub-snapshot that they were read from. The diagram at the bottom of the figure shows how the spatial hashtable file is structured.}
    \label{fig:spatialhashtable}
\end{figure}

As described in Sec.~\ref{sec:snapshots}, particles are sorted in the snapshots based on the FoF group to which they belong, with particles not assigned to any halo stored at the end. Selecting particles based on their spatial location (as necessary for \eg studies of the CGM and IGM surrounding galaxies) necessitates loading the entire coordinate list of the particles of this type, which is both wasteful and not possible on all machines due to memory constraints, while loading chunk after chunk can be very slow. In order to circumvent this issue, we provide hashtable files, created using the \texttt{sparepo}\footnote{\url{https://github.com/jborrow/sparepo}} \texttt{python} module. These files and the associated module allow users to read snapshots based upon the spatial volume requested, rather than by galaxy or halo number, in a coarse-grained way.

The structure of the files and their application to real data is demonstrated in Fig.~\ref{fig:spatialhashtable}, where we show a single `cell' from the spatial hashtable, with particles assigned to the cell represented as points (coloured by the file they were read from). The files themselves are designed to be both human and machine readable, with fast routines (accelerated with \texttt{numba}) implemented in \texttt{sparepo} to read the data. \texttt{sparepo} can additionally provide extra utility, such as reading the data with symbolic units leveraging the unyt package.

The files are structured as follows: for every snapshot there is a spatial hashtable file. Within the file, there are metadata describing the cell layout (read automatically by \texttt{sparepo}; see the documentation for a full description of these metadata), which is a Cartesian grid with 20 cells per axis (8000 cells per file), meaning each cell has a volume of roughly (5~cMpc)$^3$. This cell size was chosen to optimise both data storage volumes and reading times. For each cell in the volume, there is an independent data group named \texttt{CellX} (with \texttt{X} being a unique cell ID as described by the metadata), each containing (up to) four particle type data groups named \texttt{PartTypeP}, identical to the snapshots. Within each of these lowest level of groups, there are datasets named \texttt{FileC}, where \texttt{C} is the file chunk that contributes particles of this type to the cell. The \texttt{FileC} datasets contain a sorted array of 32 bit integers referring to the index within the chunk file (\texttt{C}) that particles residing within this chunk inhabit. This makes it possible to load the relevant chunks for a specific position (and search radius) within the volume to read both galaxies within the chunk and all associated `fluff' (i.e. the IGM). This significantly smaller array of particle properties can then be further filtered. Documentation and usage examples of these files can be found on the \texttt{sparepo} website.

\subsection{Cross-link files}
\label{sec:cross-link}
\filespecs{cross\_link.hdf5}{postprocessing/cross\_link}{all runs except the \thesanhr series}

One of the highlight features of the \thesan simulation suite is the inclusion of multiple runs with different physical models for the escape of ionizing photons and for the nature of DM. In order to enable a one-to-one comparison between objects in the different runs, we provide cross-matched catalogs linking the subhaloes at a given redshift across different runs. They were built following the same approach as the merger trees (\ie maximizing the weighted number particles shared between two subhaloes), except that instead of linking subhaloes across snapshots we link them across simulation runs. For each simulation run we provide a file named \texttt{cross\_link.hdf5}, which contains an HDF5 group for each \textit{other} run at the same resolution. Within each group, there is a dataset for each snapshot containing a list of IDs. The value in position $j$ is the subhalo global index of the object associated in the other simulation (specified by the HDF5 group) with subhalo of global index $j$ in the current simulation. More transparently, for a simulation SIM1, the subhalo of global index $j$ is paired to the one in SIM2 of global index given by the HDF5 dataset \texttt{to\_SIM2/snap\_NNN[$j$]}, where \texttt{NNN} has the usual meaning.

There is no guarantee that subhaloes are bijectively and unequivocally associated in two simulations; \ie it is possible that the subhalo $j_1$ in SIM1 is associated with subhalo $j_2$ in SIM2, but the latter is \textit{not} associated with subhalo $j_1$ in SIM1. Moreover, it is also possible that a single halo in one run is associated to multiple haloes in another one. Finally, it is possible that no association is found, in which case the cross-link will have the value -1. These cases represent a minority (typically below 5\% of the total), arising both for physical and numerical reasons, \eg due to a different power spectrum in the initial conditions, slightly different output times in different simulations, and random variability between simulations \citep[see e.g.][]{Genel2019,Keller2019,Borrow2022}.

\begin{figure}
    \centering
    \includegraphics[width=\columnwidth]{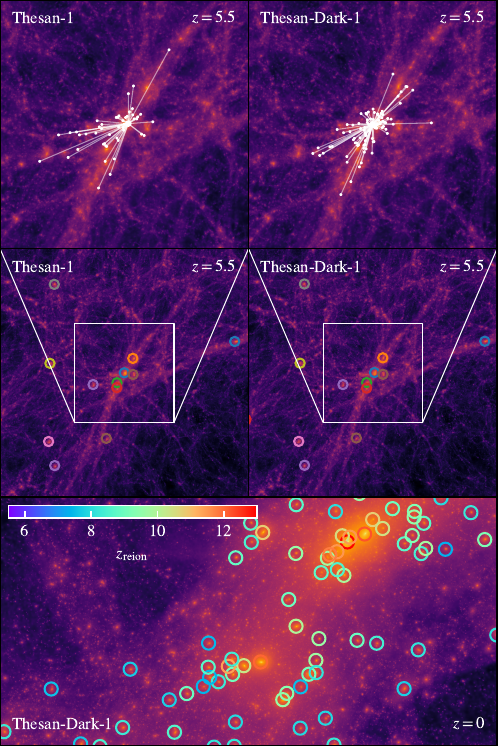}
    \caption{Example of cross-matches between \thesanone and \mbox{\thesandarkone.} The central panels show the projected dark matter density within a 10\,cMpc view of the most massive halo at $z=5.5$ in \thesanone (left) and \thesandarkone (right). Matches between all subhaloes with mass $M_{\rm H} > 10^{11}$\,M$_\odot$ within this view are shown as circles colour-matched between the two simulation volumes. To highlight potential pitfalls with the matching algorithm, the top two panels show a zoomed-in view of both simulations, with white lines showing all subhaloes that are cross-matched from the other volume to the most massive, central, subhalo. The bottom panel shows the same region, at $z=0$ in \thesandarkone, utilizing cross-links and merger trees to provide reionization timings for all dark matter subhaloes with $M_{\rm H} > 10^{11}$\,M$_\odot$.}
    \label{fig:cross_link_example}
\end{figure}

In Fig.~\ref{fig:cross_link_example} we provide an example of the cross-matching between \thesanone and \thesandarkone. The background of each panel is the projected dark matter density, with a projection depth of 10 cMpc around the central object. The central two panels show a 10 cMpc square region around the most massive halo in \thesanone, and the same matched region in \thesandarkone. All subhaloes around this object with $M_{\rm H} > 10^{11}$~M$_\odot$ are shown as cololured circles, matched between the two simulations, showing excellent agreement.

In the top two panels of Fig.~\ref{fig:cross_link_example}, we show a case where the output from the cross match files is suboptimal. Each white point and line shows a case in the sibling simulation that is matched to the central substructure, around 50 in both cases. These are subhaloes that have exchanged significant material with the central or are very low mass (and hence close to the resolution limit) and are misidentified in the alternate case as being entirely part of the central subhalo.

The bottom panel of Fig.~\ref{fig:cross_link_example} shows how the cross match catalogues can be used to provide accurate reionization timings for $z=0$ bound substructures in \thesandarkone. Objects have their reionization timing calculated within \thesanone, and are matched to \thesandarkone. They are then followed to $z=0$ using the merger trees (see Sec. \ref{sec:merger_trees}), allowing us to see the diversity of reionization timings within this highly clustered field.

\subsection{\Lya catalogs}
\label{sec:lya_catalogs}
\filespecs{Lya\_NNN.hdf5}{postprocessing/Lya}{all RMHD runs}

For all snapshots, we produce \Lya catalogs containing additional information for each halo and subhalo, saved in a single HDF5 file. These files mirror the order in the group files; \ie the additional information relative to the subhalo with global index $j$ is found in each dataset of \texttt{Lya\_NNN.hdf5} at position $j$. In particular, these catalogs provide data on continuum luminosities at $1216$, $1500$ and $2500$\,\AA, the individual ionizing luminosity contributions of stars, AGNs, collisional excitation emission and recombination, as well as the \Lya-weighted central position, velocity and velocity dispersion of the halo. Additionally, we include global \Lya properties as HDF5 attributes of the datasets. The latter are also computed for the `fuzzy' component (\ie unbound particles) within and between haloes (named `inner' and `outer' following the definition given in Sec.~\ref{sec:snapshots}), and the entire simulation box (`total').

\subsection{Lines of sight}
\label{sec:los}
\filespecs{rays\_NNN.hdf5 $\textrm{(for \texttt{N}}\geq 54$)}{postprocessing/los}{all RMHD runs except the \thesanhr series}

We extract gas quantities extracted along a set of random lines of sight (LOS) in the simulation box. The LOS have random origin and direction, have length $L_\mathrm{LOS} = 100$ cMpc and employ periodic boundary conditions when necessary. They are extracted using the \texttt{colt} code \citep[most recently described in][]{Smith+2022}, which uses the native Voronoi tessellation of the simulation, therefore retaining  the full spatial information available. For this reason, each LOS is composed of a set of segments of different length, corresponding to the intersection between the ray and one Voronoi cell. Notice that each ray has a different number of segments, determined by the native Voronoi structure. For each segment, a number of gas properties are extracted from the Voronoi cell and saved. 

We provide 150 LOS for each snapshot from the 54th onward, corresponding to $z \leq 7$. Within each LOS file, the properties extracted along the LOS are saved in a number of groups, each one containing a dataset for each ray (named simply with a progressive number, starting from 0). This means that, \eg, the density along the third ray is stored in the dataset \texttt{Density/2}. In the future we will augment the LOS data with additional relevant catalogs, \eg lightcones for cosmological integration.

\subsection{Filament catalogs}
\label{sec:filaments_catalogs}
\filespecs{filaments\_Fixed\{Mass,Number\}Tracers\_NNN.hdf5}{postprocessing/filaments}{all RMHD runs except the \thesanhr series}

We generate catalogs of cosmic filaments for all simulations within the \thesan suite. Filaments are identified in post-processing using the code \disperse \citep{Disperse1, Disperse2}. \disperse employs the Discrete Morse theory to identify cosmic filaments from the topology of the density field. In particular, filaments are defined by a set of segments connecting local maxima (`peaks') to saddle points, effectively tracing the ridges of the density field. This identification process is governed by the so-called persistence ratio ($\sigma$), a metric that quantifies the structural robustness relative to Poisson noise.
As a final step, bifurcating filaments (\ie those that connect a single critical point to multiple saddle points) are split by introducing bifurcation points to avoid double-counting of their shared part. 

\begin{figure}
    \centering
    
    \includegraphics[width=\columnwidth]{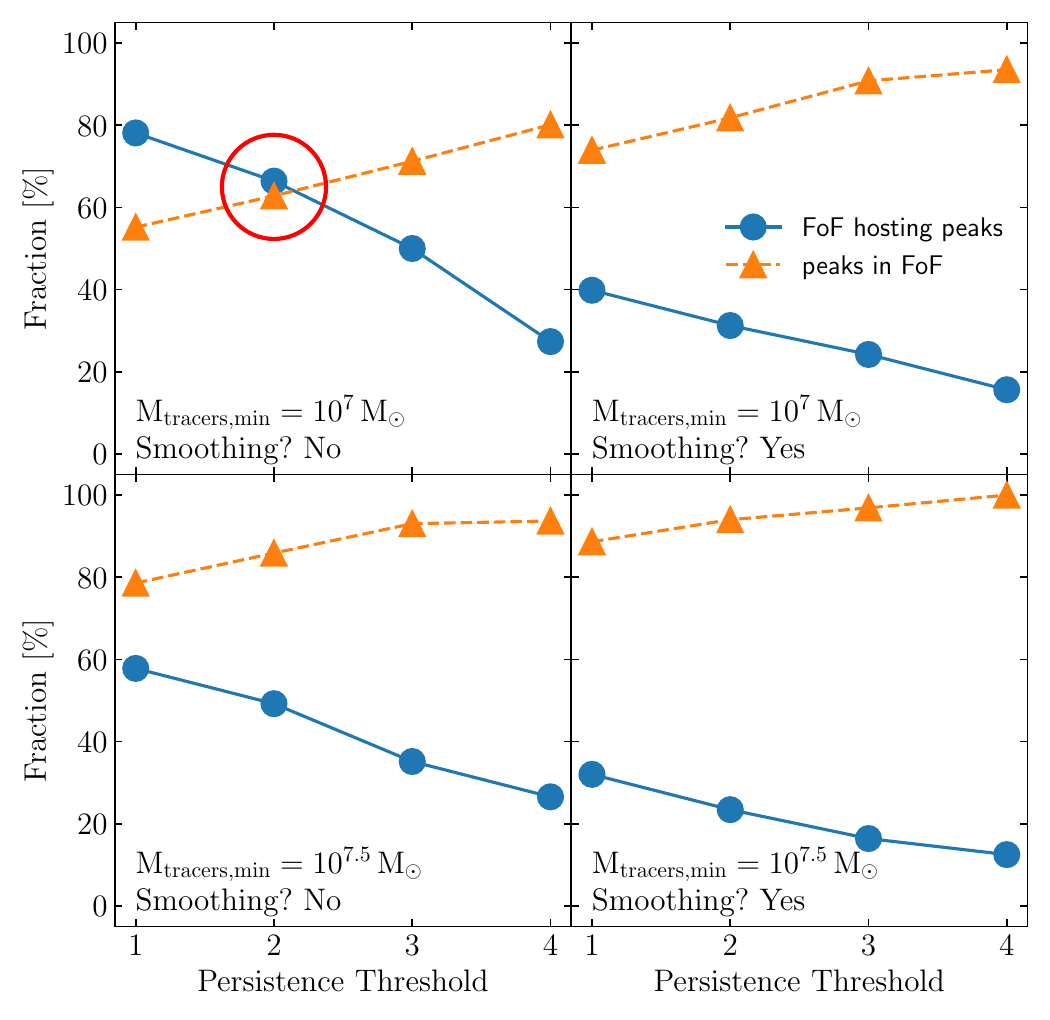}
    \caption{Summary of the calibration performed for the filament catalogs using a fixed number of tracers. The fraction of large haloes ($M_\mathrm{star} \geq 10^{11.5} \, \Msun$) hosting filament extrema (`FoF hosting peaks', solid lines) and the fraction of filament extrema within large haloes (`peaks in FoF', dashed lines) are shown as a function of \disperse parameter combinations, namely the minimum tracer stellar mass ($10^7\,\Msun$ and $10^{7.5}\,\Msun$ in the top and bottom panels, respectively), the smoothing of the reconstructed density field (applied in the right-side panels and not in the left-side ones), and the persistence threshold employed (varying along the horizontal axis within each panel). The red circle indicates the chosen parameter combination.}
    \label{fig:filaments_calibration}
\end{figure}

We have chosen to build the filament catalogs from galaxies, used as tracers of the underlying density field, to closely track the approach commonly adopted in observational studies. 
From this discrete set of tracers, \disperse reconstructs a continuous density field using the Delaunay Tessellation Field Estimator \citep{Schaap&vandeWeygaert2000}, optionally including a smoothing procedure based on the moving average over all connected vertices of the tessellation. 
We provide catalogs following two different tracer selection techniques: 
\begin{itemize}
	\item \textit{Fixed stellar mass threshold}: We use galaxies with stellar mass $M_\mathrm{star} \geq 10^{7.5} \, \Msun$, independently of the redshift considered. This results in a decreasing number of tracers with increasing redshift, as is the case in observations. The mass threshold was chosen to approximately match the smallest galaxies routinely observed by \textit{JWST} \citep[\eg][]{Santini+2023}. In this case, we smooth the inferred density field prior to filament identification and apply a fixed persistence threshold of $\sigma=3$ for all the extractions at different redshift. 
	\item \textit{Fixed number of tracers}: The previous tracer selection method results in structures characteristics of different scales being identified at each redshift, since tracers are increasingly biased with growing redshift. To counteract this effect, we follow the approach proposed by Gal\'arraga-Espinosa et al. (in prep.). Briefly, tracer selection (resulting in $N_\mathrm{tracers}$ tracers) and parameter optimization are performed at the lowest available redshift ($z=5.5$). At all other redshifts, we employ the same parameters and apply \disperse to an equal \textit{number} of tracers, selected as the $N_\mathrm{tracers}$ galaxies with highest stellar mass. The parameters explored are: the minimum mass threshold for tracers, the smoothing of the reconstructed density field, and the persistence threshold employed. Their optimization was performed by simultaneously maximizing: (i) the number of filament extrema within three virial radii of large haloes (`peaks in FoF' in Fig.~\ref{fig:filaments_calibration}), and (ii) the number of large haloes hosting filament extrema (`FoF hosting peaks' in Fig.~\ref{fig:filaments_calibration}). The first number measures the `purity' of the catalog, \ie how many spurious noise-induced filaments are identified, since these are not expected to preferentially lie close to density peaks, unlike real ones. The second measures `completeness' by estimating how many of the largest simulated objects are connected by cosmic filaments, as expected from structure formation theory. In both cases, we defined as large haloes those with $M_\mathrm{star} \geq 10^{11.5} \, \Msun$. We illustrate the results of this search in Fig.~\ref{fig:filaments_calibration}, where the fractions of `FoF hosting peaks' (solid lines) and `peaks in FoF' (dashed lines) are shown as a function of the explored parameter combinations. Coloured curves show the variation as a function of the chosen persistence threshold, while the red circle indicates the chosen parameter combination. We note that the parameter combination (7.5, N, 1.0) seems to be an even better choice (it shows a significantly higher fraction of `peaks in FoF' and similar `FoF hosting peaks'). We however decided against its use because the low persistence value produces a large number of spurious filaments, as we confirmed by investigating the filament length distribution, which showed an excess of very short filaments. 
\end{itemize}

The outputs from \disperse are re-arranged from the original format\footnote{The code used to read and re-arrange the \disperse output can be found at \url{https://github.com/EGaraldi/disperse_output_reader}.} into an \textsc{HDF5} file for ease of use, with a structure described on the \thesan website.

\subsection{Escape fraction catalogs}
\label{sec:fesc_catalogs}
\filespecs{ion\_NNN.hdf5}{postprocessing/ion}{all RMHD runs}

For all snapshots, we compute the subhalo ionizing escape fraction (\ie the fraction of ionizing photons that successfully escape the subhalo into the IGM) for each subhalo containing at least one stellar particle and four gas cells within its virial radius. These quantities are provided in catalogs directly mirroring the subhalo catalogs (\ie the quantities computed for the subhalo with global index $j$ are found in each dataset of at position $j$), contained in a single HDF5 file for each redshift. The actual calculation is fully described in Section 2 of \citet{Thesan_fesc}, and briefly summarized in the following. 

In practice, the escape fraction is defined as the ratio between the number of ionizing photons that escape the virial radius of the halo and the number of ionizing photons intrinsically emitted from all sources within the same volume. The latter includes the stellar escape fraction assumed in the simulations; \ie the fraction of photons escaping the unresolved structures around the stellar particle (see column 8 of Table~\ref{table:simulations}). We consider a photon escaped from the halo once it crosses the radius at which the average density reaches $200$ times the density of the universe. To ensure consistency, we re-compute all halo properties (mass, star-formation rate, etc.) using this definition of its boundary and provide their values in the catalogs. The intrinsic ionizing photon production rate for each star is computed employing the Binary Population and Spectral Synthesis models \citep[BPASS version 2.2.1;][]{BPASS2017, BPASS2018} with a \citet{Chabrier2003} IMF and is provided in the \Lya catalogs (Sec.~\ref{sec:lya_catalogs}). The rate of escape of ionizing photons is computed through passive Monte Carlo radiative transfer calculations performed with the Cosmic \Lya Transfer (\texttt{colt}) code \citep[most recently described in][]{Smith+2022}. Finally, whenever a stellar particle is located outside of the virial radius of any halo, we assume that the total (`halo') escape fraction equals its stellar escape fraction.

\subsection{Galaxy SED catalogs}
\label{sec:sed_catalogs}
\filespecs{SED\_NNN.hdf5}{postprocessing/SED}{\thesanone at $z=6,7,8,9,10$}

For selected redshifts equal to $z=6$, $7$, $8$, $9$, and $10$ we post-process well-resolved galaxies in the \thesanone simulation with the Monte Carlo radiative transfer code \texttt{skirt} \citep{skirt} to produce synthetic spectral energy distributions (SEDs). All galaxies have their SEDs packed in a single file by redshift. The full post-processing methodology for generating them is outlined in Section~2.2 of \citet{Thesan_lim}, and summarized in the following.

We take the gas and star distributions directly from the simulation itself, while the intrinsic nebular emission is estimated from \citet{Byler+2017}, who used photoionization calculations carried out with the \texttt{CLOUDY} code \citep{cloudy2017}. The radiation emitted by stellar particles is computed using the Flexible Stellar Population Synthesis (FSPS) model \citep{FSPS2009, FSPS2010}. The radiative transfer calculations are performed by \texttt{skirt} on a Voronoi grid constructed from positions of the gas cells in the galaxy (mirroring the structure and geometry of the AREPO data as closely as possible) and cover the wavelength range $[0.05, 200] \, \mu$m, discretized into 657 (unequal) bins. Following the considerations presented in Sec.~3.2.3 of \citet{Thesan_intro} we employ a redshift-dependent dust-to-metal ratio instead of the simulated dust distribution. Finally, we assume thermal equilibrium between the dust and the local radiation field as well as a \citet{Draine&Li2007} dust mixture of amorphous silicate and graphitic grains, including polycyclic aromatic hydrocarbons (PAHs). 

Only sufficiently well-resolved, star-forming galaxies are considered to ensure that the produced SEDs are reasonably converged. In particular, an SED is computed for a galaxy if: (i) the stellar mass within twice the stellar half-mass radius is greater than 50 times the baryonic mass resolution of the simulation, and (ii) it contains at least one stellar particle younger than 5 Myr (at the resolution of the simulation this roughly translates to haloes with a minimum star-formation rate of just below $0.1 \, \Msun$~yr$^{-1}$).

\subsection{Initial conditions}
\label{sec:ics}
We release the initial conditions (ICs) used for all the \thesan runs. For the original \thesan runs, all ICs correspond to the same (95.5~cMpc)$^3$ volume of the Universe at $z=49$, sampled at two different resolutions (corresponding to \thesanone and \thesantwo, respectively). For the \thesanhr series, the ICs correspond to the same (but different from the original \thesan) (5.9~cMpc)$^3$ volume of the Universe at $z=49$, sampled assuming different models for the dark sector, plus a set of ICs of the same volume at 8 times higher mass resolution (corresponding to the \thesanvhr run) and a set covering a larger (11.8~cMpc)$^3$ box (for the \thesanhrlarge) run. 
All ICs are generated (and provided) as DM-only files. Baryons were included by splitting particles (according to the baryon-to-DM-density ratio) at the startup of the simulation. All the ICs were generated using the N-GenIC version contained in the \gadgetfour code \citep{Springel2021}, using second-order Lagrangian perturbation theory and fixing the amplitude of each density perturbation mode to the power spectrum expectation value \citep[\textit{fixed} approach from][]{AnguloPontzen2016}. This ensures that the ICs generated are fully representative of the average Universe on all scales sampled.

\section{Usage considerations}
\label{sec:usage}

\subsection{Data retrieval}
\label{sec:data_retrieval}
\begin{figure}
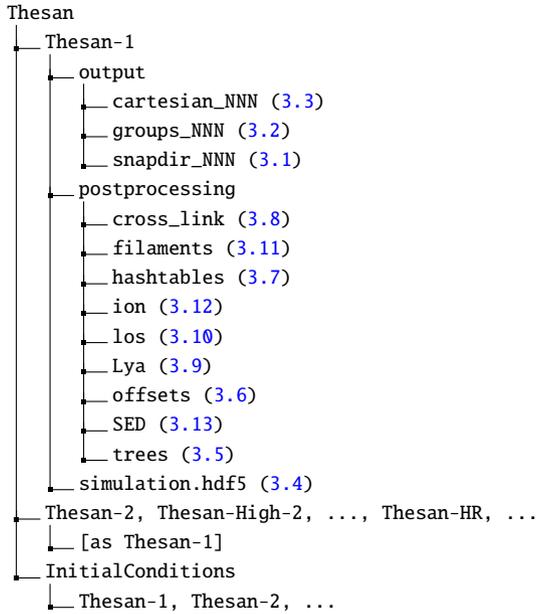

\dirtree{%
.1 Thesan.
.2 Thesan-1.
.3 output.
.4 cartesian\_NNN (\hyperref[sec:cartesian]{3.3}).
.4 groups\_NNN (\hyperref[sec:fof_catalogs]{3.2}).
.4 snapdir\_NNN (\hyperref[sec:snapshots]{3.1}).
.3 postprocessing.
.4 cross\_link (\hyperref[sec:cross-link]{3.8}).
.4 filaments (\hyperref[sec:filaments_catalogs]{3.11}).
.4 hashtables (\hyperref[sec:hashtables]{3.7}).
.4 ion (\hyperref[sec:fesc_catalogs]{3.12}).
.4 los (\hyperref[sec:los]{3.10}).
.4 Lya (\hyperref[sec:lya_catalogs]{3.9}).
.4 offsets (\hyperref[sec:offsets]{3.6}).
.4 SED (\hyperref[sec:sed_catalogs]{3.13}).
.4 trees (\hyperref[sec:merger_trees]{3.5}).
.3 simulation.hdf5 (\hyperref[sec:simulationhdf5]{3.4}).
.2 Thesan-2, Thesan-High-2, ..., Thesan-HR, ....
.3 [as Thesan-1].
.2 InitialConditions. 
.3 Thesan-1, Thesan-2, .... 
}
    \caption{Organization of the public \thesan data directory with subsection references. For the sake of brevity, we have omitted from the figure some of the less important files, as well as the content of the \texttt{postprocessing} sub-directories, which have been thoroughly described in Sec.~\ref{sec:data_products} or the online documentation.}
    \label{fig:toc}
\end{figure}

We rely on the Globus\footnote{\url{https://www.globus.org/}} file transfer service to make the \thesan data public. This infrastructure provides a solid, flexible and efficient system to transfer data, alongside with fine-grained customization options. It features a browser-based interface as well as command-line and python access options, automatic verification of data integrity and autonomous handling of connection interruptions. The Globus workflow involves \textit{endpoints}, \ie machines where the Globus application has been installed. Users can initialize data transfer between two endpoints and let Globus handle the execution. Alternatively, it is possible to download the data via simple direct download at the price of losing the features listed above. The \thesan data are stored in a collection named `Thesan', stored at the endpoint `MPCDF DataHub Server'. A direct link to the collection can be found at \url{https://www.thesan-project.com/data.html}, alongside detailed documentation and a short Globus tutorial. The organization of the released data into directories is shown in Fig.~\ref{fig:toc}. \rev{}{Please notice the citation policy on the website, which ensure that the contributions of all individuals involved in the development of the \thesan simulation suite are properly acknowledge.}

\subsection{Physical considerations}
\label{sec:physical_considerations}
\thesan builds upon the successful IllustrisTNG galaxy formation model, inheriting both its strengths and limitations. Notably the inclusion of self-consistent radiation transport makes \thesan particularly well suited for studying the $z\gtrsim5$ Universe. In the following, we outline and briefly comment on several aspects of \thesan that warrant cautious interpretation.

\subsubsection{Numerical convergence}

\begin{figure}
    \centering
    \includegraphics[width=\columnwidth]{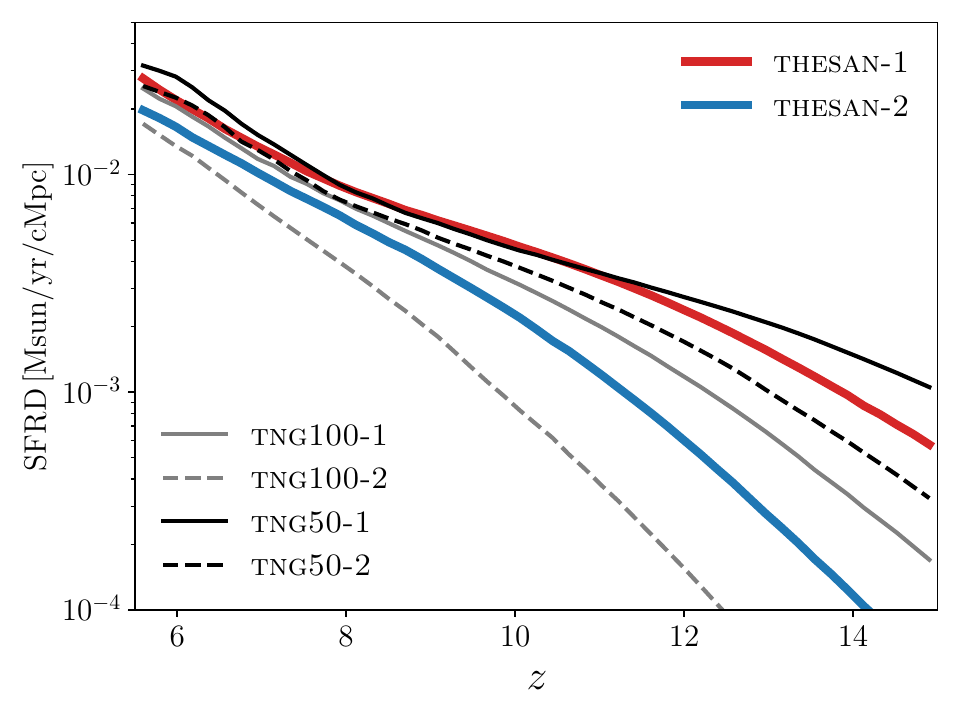}
    \caption{Comparison of the star-formation rate density in the \thesanone, \thesantwo, IllustrisTNG-100-1, IllustrisTNG-100-2, IllustrisTNG-50-1, and IllustrisTNG-50-2 simulations. As discussed in \citet{ThesanHR} the inclusion of radiation in \thesan leads to a suppression of star formation relative to models using a spatially-uniform UV background at $z \gtrsim 10$.}
    \label{fig:sfrd_comparison}
\end{figure}

Numerical convergence is a crucial factor in assessing the reliability of simulation outcomes. Different physical quantities often reach convergence at different resolution levels, and careful comparisons are required to understand the quantitative accuracy of measurements derived from the simulations.

For \thesan it is critical to consider the star-formation rate (SFR) density (SFRD) and its convergence, due to the direct link between the SFRD and ionizing photon budget, especially given our fixed choice of $\fesc$ for stars. In Fig.~\ref{fig:sfrd_comparison} we show the SFRD for both \thesanone and \thesantwo, and highlight that \thesanone exhibits a higher star-formation rate at high redshifts --- a factor of $\approx 1$ dex at $z \gtrsim 13$, decreasing to a factor of $1.5$ at $z \lesssim 8$. With a mass resolution eight times higher than that of \thesantwo, along with correspondingly smaller softening scales, \thesanone is capable of resolving higher gas densities and hence earlier star formation.

To show that we have almost reached numerical convergence of the SFRD in \thesanone, we can employ the already existing IllustrisTNG-100 and IllustrisTNG-50 simulations, which have box volumes with side lengths of 106.5\,cMpc and 51.7\,cMpc, respectively. Fig.~\ref{fig:sfrd_comparison} shows four comparison simulations in these two box volumes with the following mass resolutions: TNG100-1 ($m_{\rm gas}=1.4\times10^6$~M$_\odot$), TNG100-2 ($m_{\rm gas}=1.1\times10^7$~M$_\odot$), TNG50-1 ($m_{\rm gas}=8.4\times10^4$~M$_\odot$), and TNG50-2 ($m_{\rm gas}=6.8\times10^5$~M$_\odot$). \thesanone has a resolution close to TNG50-2, and \thesantwo has a resolution between TNG100-1 and TNG100-2. There are only relatively minor differences between the SFRD in TNG50-1, TNG50-2, and \thesanone, indicating that we are close to a numerically converged model. Specifically, the SFRD differences above $z\gtrsim14$ are within a factor of $\approx 2$, with the onset of star formation only occurring around $z \approx 16$. For \thesantwo, however, there are significant differences between the two TNG100 resolution levels, with \thesantwo lying in between, indicating that numerical convergence has not yet been attained for this simulation.

\rev{}{A second key numerical parameter worth discussing here in more details is the minimum simulation volume required to provide converged predictions of reionization properties, which remains unclear. In particular, \citet{Iliev+2014} found that volumes of approximately 100 $\hMpc$ are sufficient for good convergence of most simulated IGM properties, with the exception of the 21cm signal requiring volumes of at least 250 $\hMpc$. However, these simulations do not account for Lyman-limit system, resulting in a likely over-estimation of the necessary volumes \citep[see also the discussion in section 6.4 of][]{GnedinMadau_review}. Finally, \citet{Gnedin+2014} showed that in the CROC simulations (in many ways similar to \thesan in terms of physical richness, resolution and ability to reproduce observed IGM properties) boxes as small as 20 $\hMpc$ are sufficient for the convergence of the ionised bubble size distribution at $z>7$. Concerning \thesan, we do not have access to runs with equal resolution in different box sizes, but figure 2 of \citet{Thesan_bubbles} shows no signs of truncation of the bubble size distribution until towards the very end of the reionization history. Moreover, we have confirmed that the simulated reionization history, IGM temperature, CMB optical depth and bubble size distribution are practically unchanged when the entire simulation box is divided into 8 independent sub-boxes (see figure 3 of \paperII for a visualisation of the first three). While not exactly equivalent to a volume convergence study, these facts  give us confidence that that reionization properties are likely converged with respect to the simulated volume in our runs.}

\subsubsection{Reionization history}
As described in Sec.~\ref{sec:simulations}, the escape fraction of ionizing photons from stars was tuned to obtain a model within which reionization occurs late, with this calibration performed at the \thesantwo resolution. Results from \paperII suggest that reionization in \thesanone may still end slightly too early. This discrepancy is attributed to a resolution-induced factor of $\sim 3$ excess in the early star-formation rate density, as seen in Fig. \ref{fig:sfrd_comparison}. Given that $\fesc$ was not re-calibrated between different resolution levels, \thesanone consequently has a higher budget of ionizing photons.

\subsubsection{Cosmic dust}
We advise the reader to exercise caution regarding the cosmic dust content of massive galaxies within the \thesan simulation suite. Preliminary analysis suggests that the dust buildup may be too low compared to observational data. The main evidence is the insufficient attenuation at the bright end of the UV luminosity function when the simulated dust content is used (see Fig.~11 of \paperI). While this discrepancy is not entirely surprising, since we made the deliberate choice to not tweak any parameter in the model and adopt the original calibration by \citet{McKinnon+16, McKinnon+17}, which is based on Milky Way data. However, this evidence is not conclusive, since the issue may lie in the attenuation law used for dust. We are currently investigating this topic further.

\subsubsection{Unresolved interstellar medium}
\thesan employs the two-phase interstellar medium (ISM) model proposed by \citet{Springel&Hernquist03}. By design, this model precludes any attempt to self-consistently resolve the multi-phase ISM. This approach is necessary to achieve the large volume of the \thesan runs, which prevents us from reaching the resolution required for more self-consistent treatments of the ISM. Consequently, the cold dense and molecular phases of the ISM, giant molecular clouds, and individual stellar birth clouds are not explicitly resolved. The modelling of observables predominantly arising from these unresolved phases should be undertaken with care.

\subsection{Data analysis}
The majority of the data released, including snapshots, group catalogs, merger trees, and offset files, have structures similar to that of the IllustrisTNG datasets. This facilitates the straightforward adaptation of existing tools for use with \thesan data. To further streamline data utilization associated with this paper, we have updated the widely-used \texttt{python} module \texttt{illustris\_python} (along with its twin modules for the \texttt{julia}, \texttt{MATLAB}, and \texttt{IDL} programming languages) to work with \thesan data and to support the loading of our Cartesian outputs\footnote{Using this updated version requires re-downloading and re-installing the packages from: 
\begin{itemize}
    \item[] \hspace{-0.3cm} \texttt{python}: \url{https://github.com/illustristng/illustris_python}
    \item[] \hspace{-0.3cm} \texttt{julia}: \url{https://github.com/illustristng/illustris_julia}
    \item[] \hspace{-0.3cm} \texttt{MATLAB}: \url{https://github.com/illustristng/illustris_matlab}
    \item[] \hspace{-0.3cm} \texttt{IDL}: \url{https://github.com/illustristng/illustris_idl}
\end{itemize}}.

\section{Comparison with JWST results}
\label{sec:jwst_comparison}
The successful launch and deployment of the \textit{JWST} has quickly transformed the study of 
the $z\gtrsim5$ Universe. Within just a year, it has already started to reshape our understanding of primeval galaxy formation and reionization. 
Simultaneously, it has significantly extended our ability to test the current models of galaxy formation in the young Universe. This is of crucial importance as these models were typically devised and calibrated when such observations were not available, and therefore provide genuine predictions about the properties of the galaxy candidates recently uncovered. In fact, many inherently different galaxy formation models have reached comparable success in reproducing the observed galaxy population, both at low \citep{TNG_Marinacci, TNG_Naiman, TNG_Nelson, TNG_Springel, TNG_Pillepich, magneticum, romulus25, eagle, newhorizon, massive-black-ii} and high redshift \citep{SPHINX, Thesan_intro, Thesan_igm, Thesan_Lya, CROC, Obelisk,Aurora}. 
Comparing their predictions to new data in a redshift range relatively uncharted so far is therefore a key test of these models, one that could differentiate these models, and thus a prime way to advance our knowledge of the onset of galaxy formation. 

In this section we compare the predictions of \thesan with a selection of results from the \textit{JWST} to serve multiple objectives. First, this exercise tests the reliability of the \thesan simulation suite, pinpointing areas for refinement. Second, contrasting these results with similar analyses performed on simulations employing different physical and numerical prescriptions highlights the strengths and weaknesses of each model. This, in turn, improves our 
understanding of primeval galaxy formation and eventually guides the development of the next generation of numerical simulations. Third, these comparisons serve as examples of how the publicly released \thesan data can be leveraged for the interpretation of observations. To facilitate the latter, we specify at the beginning of each subsection the data products employed for the ensuing analyses and discussions.

\begin{figure*}
    \centering
    \includegraphics[width=\textwidth]{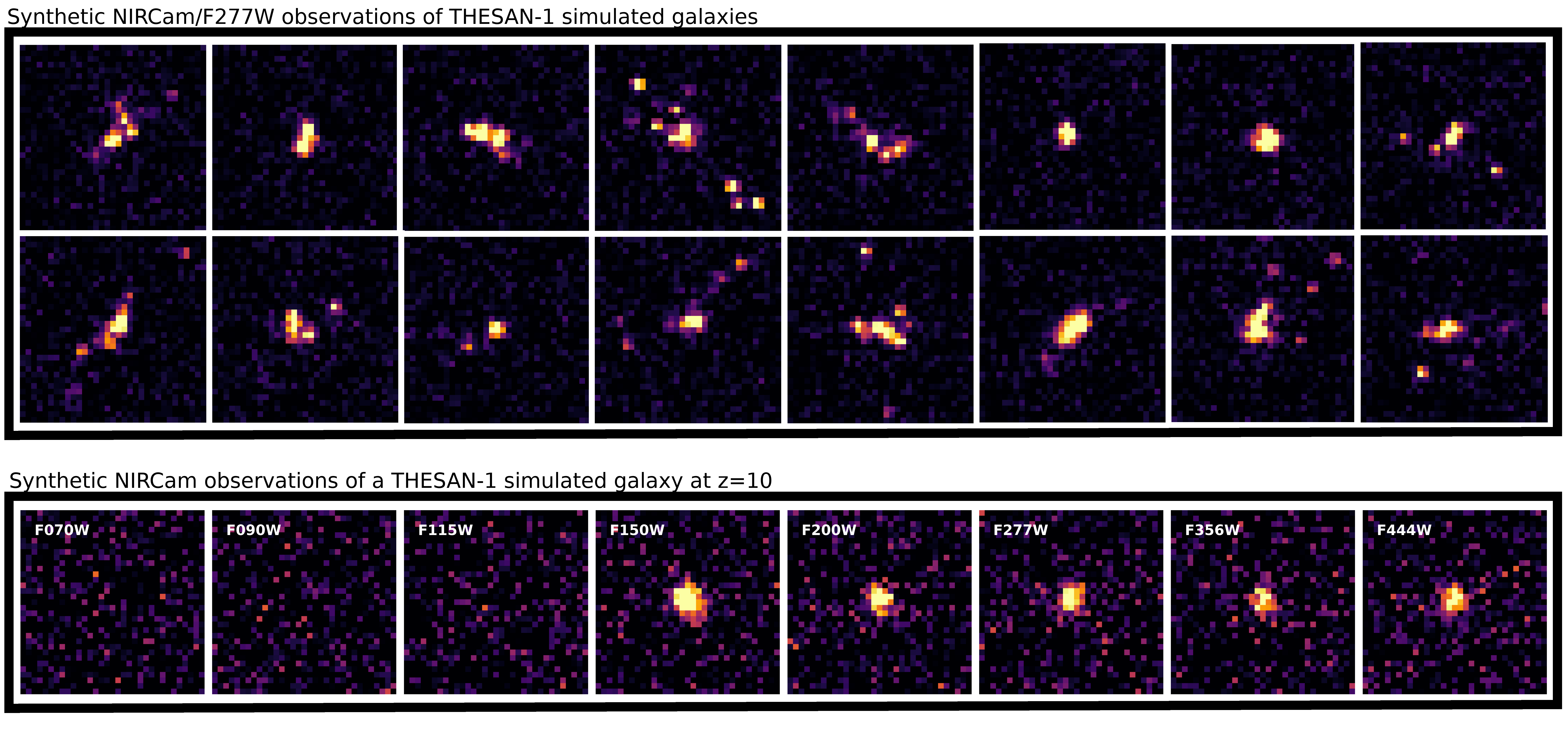}

    \caption{\textit{JWST} mock images of bright galaxies (top two rows) found in \thesanone at $z=10$, using the \texttt{F277W} filter and a noise level comparable to the early science release data in the field of SMACS J0723.3-7327. The bottom row shows how the largest galaxy in \thesanone at $z=10$ would appear in a series of \textit{JWST} filters (reported in the top left corner of each plot).}
    \label{fig:nircam_mocks}
\end{figure*}

\subsection{Galaxy spectra and images}
\dataused{snapshots, offset files}

Among the first and most surprising results from the \textit{JWST} is the identification of a significant number of galaxy candidates at $z\gtrsim8$ with large stellar masses \citep[\eg][]{Atek+2023, Castellano+2022, Labbe+2022, Naidu+2022, Tacchella+2022, Austin+2023, Curtis-Lake+2023, ceersI, Morishita+2023, Yan+2023}. 
While a thorough comparison with comparison with each one of these (and many more) works is beyond the scope of this paper, we provide a methodological example using individual galaxies extracted from the \thesanone simulation. Utilizing the \texttt{skirt} code \citep{skirt}, we simulate the propagation of radiation until it escapes the galaxy \citep[for a detailed description of the SED production, see][]{Thesan_lim}. We additionally account for attenuation from the intervening intergalactic medium (IGM) following \citet{Madau+95}. Finally, we coarsen the images to match the NIRCam resolution of $0.07$ arcseconds, corresponding to approximately $0.3$ kpc at $z=10$, and add Gaussian random noise assuming a signal-to-noise of $30$ for the image, comparable to the one achieved for $z\gtrsim8$ galaxies in the field of SMACS J0723.3-7327 \citep{JWST_ERO}. The results of this procedure are presented in Fig.~\ref{fig:nircam_mocks} for a selection of the brightest \thesanone galaxies at $z=10$. The mock images shown in the top two rows correspond to the view of the galaxies in the \texttt{f277w} filter of the \textit{JWST}, while the bottom row shows the synthetic observation of one galaxy across a number of \textit{JWST} filters (reported in the top left corner of each plot). All panels have a linear size of $10$\,kpc (33 pixels). 
Even from such a small sample it is possible to appreciate the diverse variety of galactic morphologies, ranging from isolated galaxies to active mergers. In particular, we also observe aligned structures, reminiscent of the cosmic filaments observed by \citet{Wang+2023_filament}. We have not yet investigated whether these are physically-associated structures or merely an effect of galaxy selection and alignment. The released \thesan data are well-suited for such investigations.

Encouraged by this qualitative agreement with recent observations, we proceed to a more quantitative exploration in the following subsections.

\subsection{The galaxy main sequence at $z>6$}
\dataused{group catalogs, SEDs}

\begin{figure}
    \centering
    \includegraphics[width=\columnwidth]{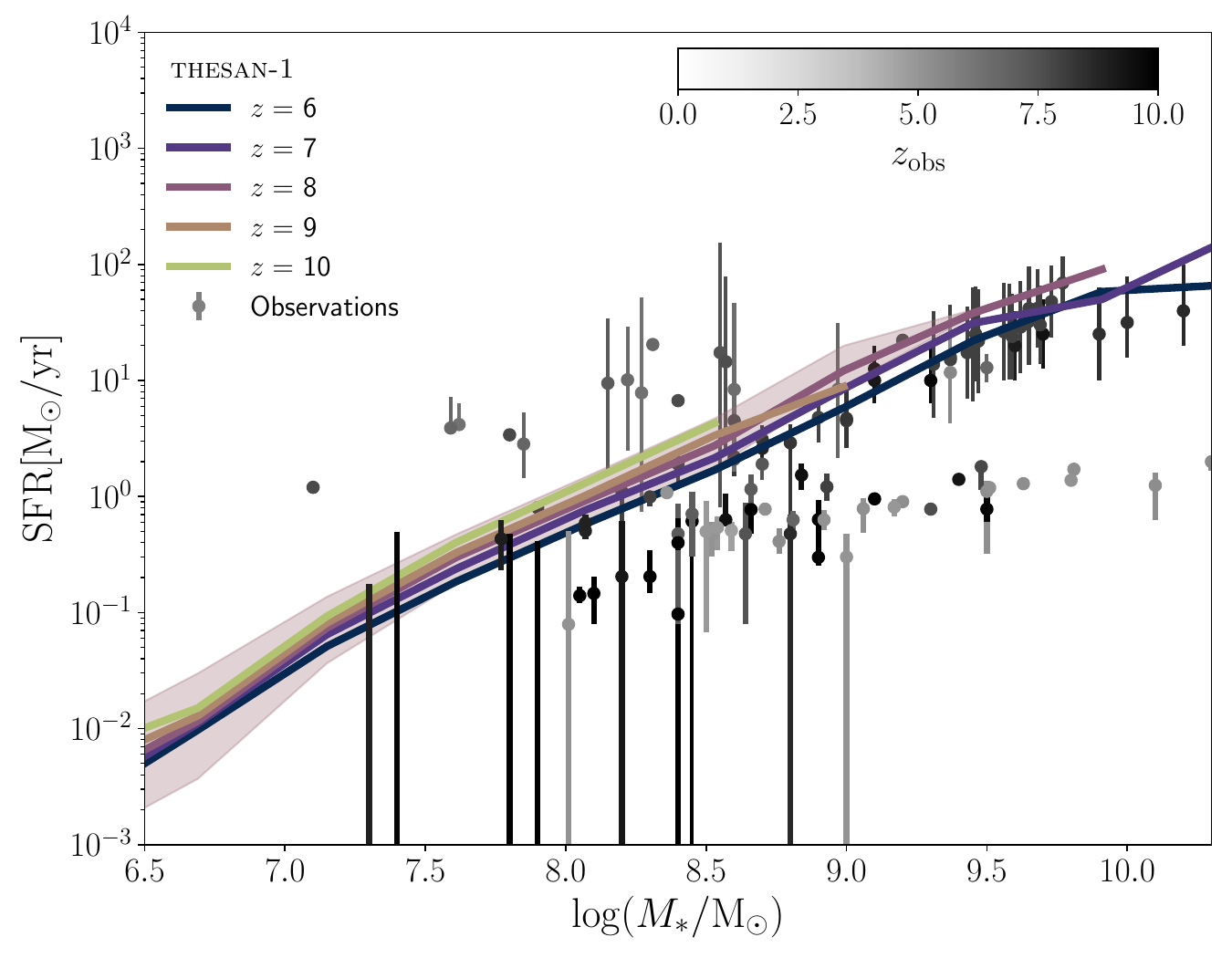}
    \caption{Galaxy main sequence at $z=6$, $7$, $8$, $9$, and $10$ in \thesanone (coloured curves), compared to the values from observations by  \citet{Leethochawalit+2023,Tacchella+2022,Stefanon+2022,Roberts-Borsani+2022,Finkelstein+2022,Chen+2023,Fujimoto+23,Arrabal-Haro+23a,Arrabal-Haro+23b,Long+2023,Jung+23,Atek+23,Sun+23,Treu+23,Asada+23,Perez-Gonalez+23}. All observations are colour-coded by their redshift, as reported in the top colour bar. \rev{}{The shaded region shows the central 68\% of data, only for $z=8$ for visual clarity.}}
    \label{fig:GMS-z}
\end{figure}

The majority of low-redshift galaxies adhere to a well-established relationship between their total stellar mass ($M_\mathrm{star}$) and their star-formation rate, commonly referred to as the galaxy main sequence (GMS). Whether such a relation holds at high redshifts is currently being investigated by the \textit{JWST} \citep[\eg][]{Leethochawalit+2023, Tacchella+2022, Stefanon+2022, Roberts-Borsani+2022, Finkelstein+2022, Chen+2023}. In Fig.~\ref{fig:GMS-z}, we present this relation for \thesanone at integer redshifts between 6 and 10, depicted as solid lines. For the sake of clarity, the central 68 per cent of the data is shown only for $z=8$ \rev{}{as a shaded region}, although the distribution is similar for the other redshifts. The figure also reports a number of recent observations about candidate galaxies identified via the \textit{JWST}. We find an overall good agreement between these results and our model predictions, with two notable exceptions: (i) For galaxies with stellar masses $M_\mathrm{star} \lesssim 10^{8} \, \Msun$, \citet{Stefanon+2022} and \citet{Chen+2023} report significantly higher SFRs than those found in \thesanone. (ii) For galaxies with stellar masses $M_\mathrm{star} \gtrsim 10^{9} \, \Msun$, the data appear to bifurcate into two distinct sequences, one in excellent agreement with \thesanone and a second with galaxies exhibiting SFR $50$--$100$ times lower. The frequency and origins of these quenched galaxies remain topics of ongoing debate, especially as only the advent of \textit{JWST} uncovered such population of galaxies. Below we discuss a recent extreme example of such a quenched galaxy, as reported by \citet{Looser+2023}.

\begin{figure}
    \centering
    \includegraphics[width=\columnwidth]{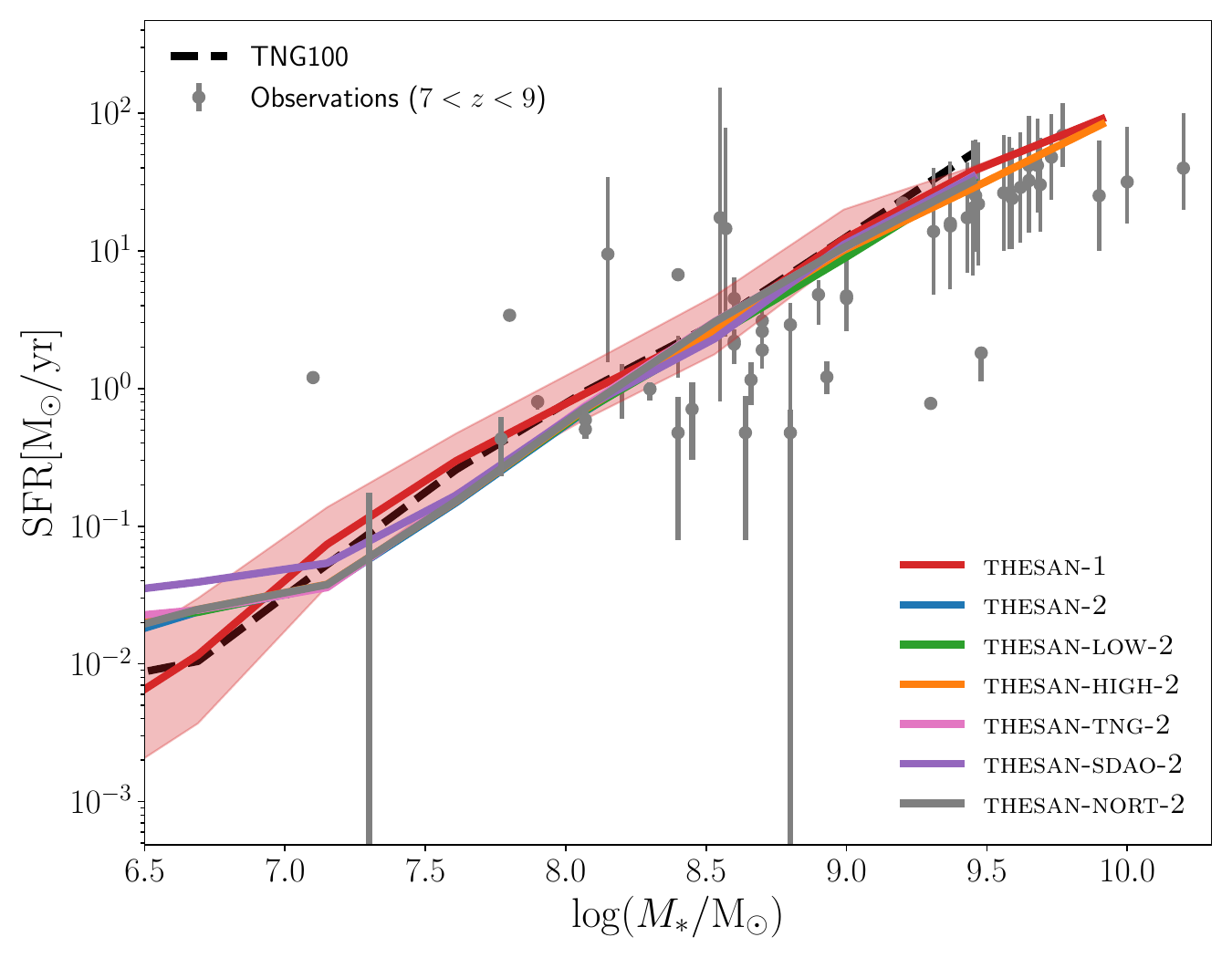}
    \caption{Same as Fig.~\ref{fig:GMS-z} but showing only the relation at $z=8$ for the different \thesan runs as well as for TNG100 and for observations at $7 \leq z \leq 9$. \rev{}{In this case, the shaded region showing the central 68\% of data is shown only for $z=8$ (again for visual clarity).}}
    \label{fig:GMS-all}
\end{figure}

\thesan predicts a modest decrease of the normalization of the GMS over cosmic time, almost uniformly across the mass range covered by the simulation. Utilizing the different runs within the \thesan suite, we can predict the influence of different physical models on the GMS. This is shown in Fig.~\ref{fig:GMS-all}, where we present the GMS at $z=8$ for the different \thesan runs. Over the mass range covered here, there is no apparent difference in the GMS between the different runs. The only minor effect is a small suppression of star formation at the lowest stellar masses in the \thesantwo runs, dictated by the limited resolution within the smallest haloes. However, as demonstrated in \citet{ThesanHR_noncdm} using the \thesanhr simulation set, different DM models do have an impact on galaxies at lower mass scales.

\subsubsection{Quenched galaxies in \thesan}
Recently, \citet{Looser+2023} reported the discovery of a galaxy with a stellar mass of $M_\mathrm{star} \approx 10^{8.7} \, \Msun$ and an exceptionally low star-formation rate ($\log({\rm SFR} [\Msun/\mathrm{yr}]) \approx -2.5$). The SED modelling of this galaxy indicates that this object was quenched between $10$ and $38$\,Myr before it was observed, after a starburst lasting $\sim50$ Myr. To investigate the presence of similar objects in \thesan and explore the conditions under which they might occur, we recompute for each galaxy its SFR over a time window of $\Delta t = 38$\,Myr. 

The \thesanone simulation does not contain any galaxies exactly matching all of these quenching properties, but there is an analogous population of reasonably similar objects with $\log(\mathrm{SFR}) \sim -2$. In fact, there are at least two different reasons why \thesanone is not expected to exactly match such observations, namely: (i) the limited volume restricts the range of galaxy properties that can be represented and (ii) the finite baryonic resolution limits the minimum SFR that can be resolved. In fact, the lowest SFR over a time window $\Delta t$ that can be obtained is simply the minimum stellar mass that can be resolved divided by $\Delta t$ itself. Given the gas mass resolution of \thesanone (see Table~\ref{table:simulations}), and the fact that \arepo is designed to keep gas masses within a factor of $2$ of the target resolution $m_\mathrm{gas}$, we can estimate the smallest SFR detectable in \thesanone as:
\begin{equation}
    {\rm SFR}_\mathrm{min} = 0.5 \, m_\mathrm{gas} / \Delta t \sim 10^{-2} \, \Msun/\mathrm{yr} \, ,
\end{equation}
where we have used $\Delta t = 38$\,Myr. Therefore, \thesanone is not able to detect an SFR as low as the one reported by \citet{Looser+2023}. In addition, we have disregarded any potential obscuration effects from cosmic dust 
which could further conceal the light from newborn stars. 
\rev{}{Notice that this calculation does not apply to the SFR reported \eg in Figs. \ref{fig:GMS-z} and \ref{fig:GMS-all}, since in these cases we report the instantaneous SFR,  computed internally by the simulation at run time from the gas bulk properties only \citep[following][]{Springel&Hernquist03}, and is therefore not affected by mass resolution.}

\rev{}{In \citet{Thesan_sizes}, we have expanded on this analysis using a more flexible definition of quenching. This resulted in the identification and characterisation of high-mass quenched objects.}

\subsection{Obscured star formation}
\dataused{group catalogs, SEDs}

The phenomenon of obscured star formation is a crucial consideration in the study of galaxy evolution. Newborn stars typically emit intense UV radiation, rendering the UV luminosity a useful proxy for the star-formation activity of a galaxy. However, when the galaxy hosts a substantial dust content, the latter can absorb and reprocess a significant fraction of UV light into the infrared (IR), effectively hiding it. This is often referred to as obscured star formation, and can be revealed by detecting the IR light emitted by cosmic dust \citep{Madau&Dickinson2014}. Recent observational efforts have started to constrain the fraction of obscured star formation in the Universe across cosmic time, yielding precious information on both the star-formation activity and the dust content of ancient galaxies.

\begin{figure}
    \centering
    \includegraphics[width=\columnwidth]{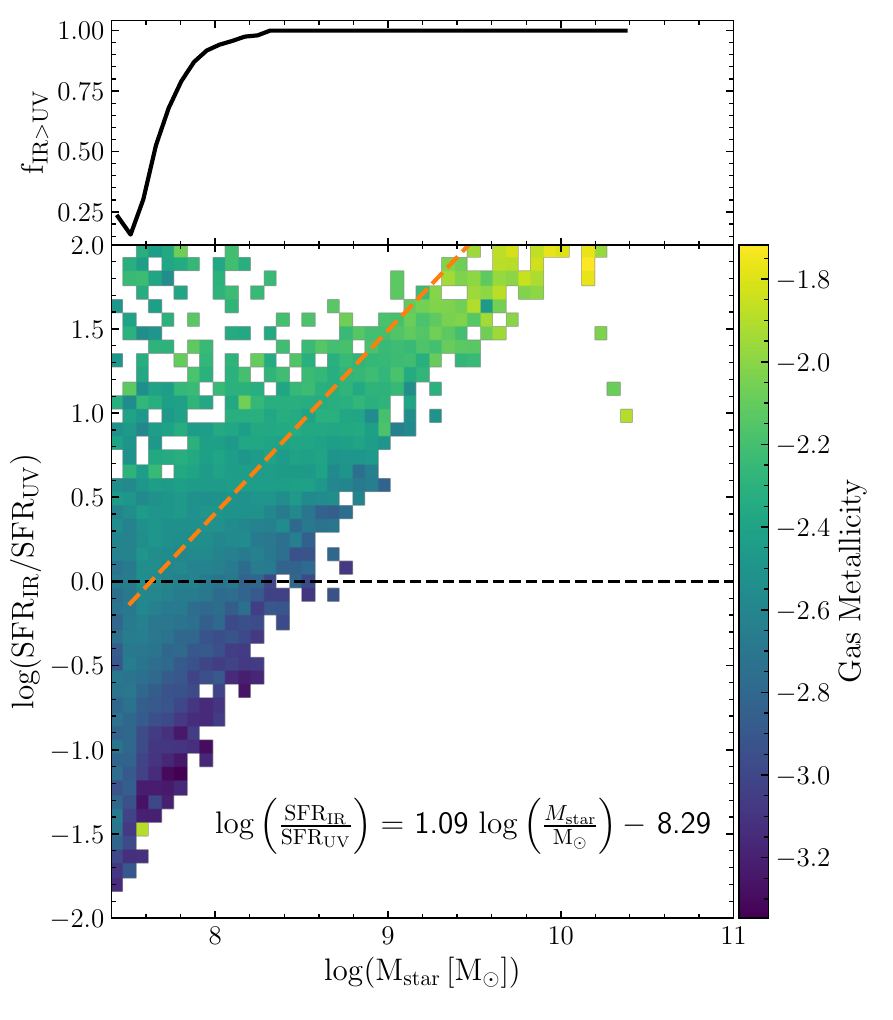}
    \caption{\textit{Top:} Fraction of obscured star formation in simulated galaxies as a function of their stellar mass in \thesanone. \textit{Bottom:} Distribution of the ratio between the IR- and UV-inferred SFR at $z=6$ for star-forming galaxies in \thesanone and their stellar mass, colour-coded by the average gas metallicity of the galaxies in each pixel, as a proxy for their dust content. Obscured star formation dominates in more massive galaxies, where the largest reservoirs of cosmic dust (reflected in their gas metallicity) are found, while it is subdominant in smaller objects. At fixed stellar mass, the relevance of obscured star formation correlates with the gas metal content. The orange dashed line shows the best-fit linear relation, also reported in the bottom right.}
    \label{fig:SFR_IR_UV}
\end{figure}

Thanks to the rich physics contained in the \thesan simulations, it is possible to estimate the amount of obscured and unobscured star formation for each galaxy, albeit with the limitations described in Sec.~\ref{sec:physical_considerations}. In order to properly compare our results to observations, we follow \citet{Shen+2022} and calculate for each galaxy obscured and unobscured SFRs from their synthetic SED (see Sec.~\ref{sec:sed_catalogs}). We convert the total luminosity in the UV ($1450 \AA$ to $1550 \AA$) and IR (8 $\mu$m to 1000 $\mu$m) bands directly into SFRs using the conversion rates provided by \citet{Madau&Dickinson2014}\footnote{We have modified their UV-luminosity-to-SFR conversion rate following the prescription of \citet{Driver+2013} to account for the different IMFs assumed in \citet{Madau&Dickinson2014} \citep[who follow][]{SalpeterIMF} and \thesan \citep[which uses][]{Chabrier2003}.}. In Fig.~\ref{fig:SFR_IR_UV}, we show the ratio between the IR- and UV-inferred SFRs at $z=6$ for star-forming galaxies in \thesanone, plotted as a function of their stellar mass. Each pixel of this two-dimensional map is colour-coded according to the average gas metallicity of the galaxies it represents, reflecting their cosmic dust content. As expected, more massive galaxies are dominated by obscured star formation, thanks to the large amount of dust that they host.  In contrast, low-mass galaxies are well characterized by UV-inferred star formation. The top panel of the figure shows the average fraction of obscured star formation in galaxies as a function of their stellar mass.

As a simple example of how this knowledge can help decipher observations, we provide the best-fit linear relation between the IR-to-UV SFR ratio and $\log(M_\mathrm{star} / \mathrm{M}_\odot)$ for the \thesanone galaxies at $z=6$. This can provide useful information to gauge the amount of star formation revealed or obscured in a given observation. We caution that this approach is simplistic and is primarily intended as an example of the science enabled by \thesan. However, the IR-to-UV SFR ratio depends on the galaxy gas metallicity, and potentially more properties, that should be taken into account. 

\subsection{The mass-metallicity relation at $z>6$}
\dataused{group catalogs}

\begin{figure}
    \centering
    \includegraphics[width=\columnwidth]{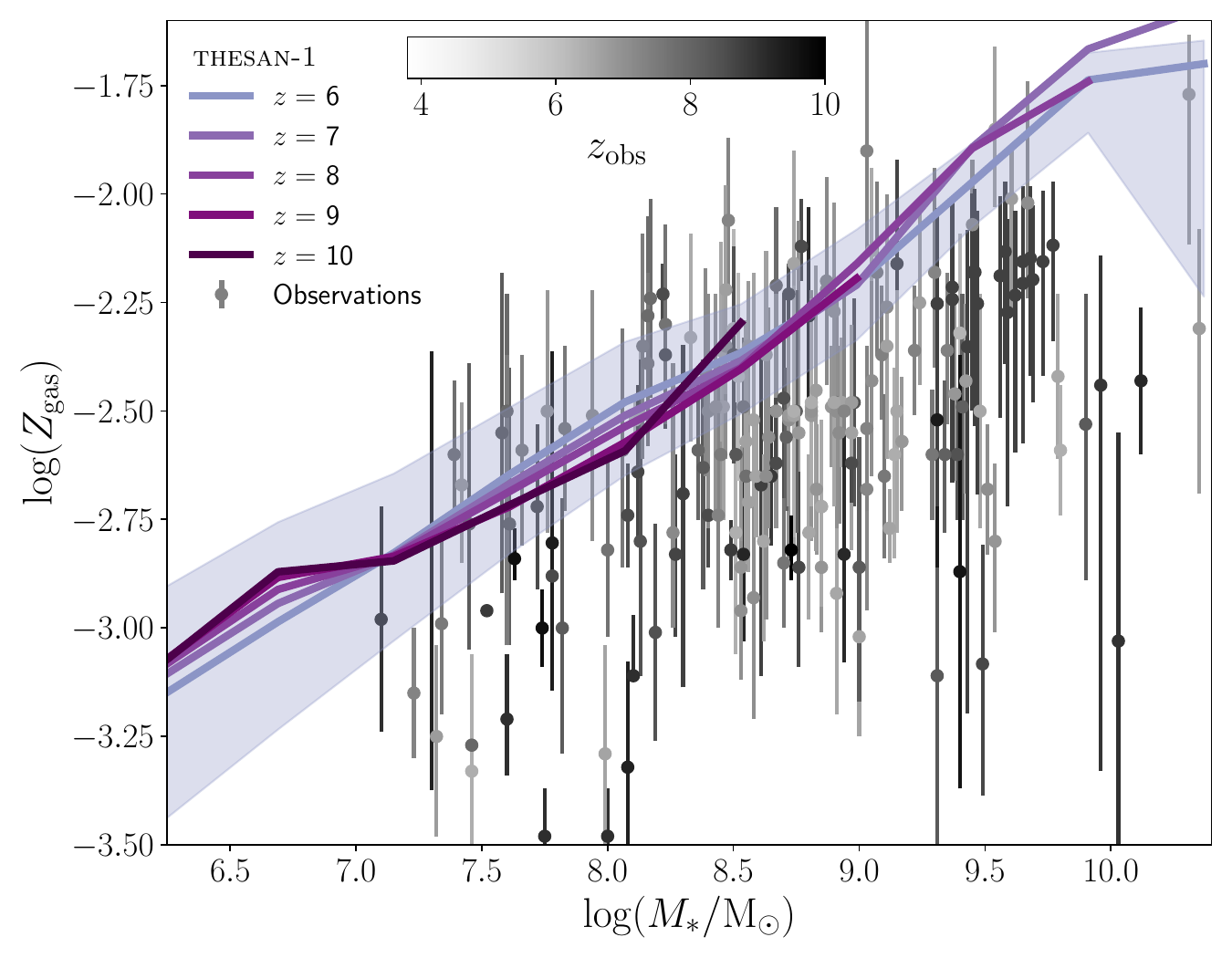}
    \caption{Mass-metallicity relation at $z=6$, $7$, $8$, $9$, and $10$ in \thesanone (coloured curves), compared to observations over the redshift range $z \gtrsim 4.8$ from \citet{Trump+2023,Langeroodi+22,Nakajima+23,Schaerer+22,Bunker+23,Boyett+23,Fujimoto+23,Curti+23a,Roberts-Borsani+22,Williams+23}, colour-coded by their redshift as indicated in the colourbar at the top.}
    
    \label{fig:MZR-z}
\end{figure}

\begin{figure}
    \centering
    \includegraphics[width=\columnwidth]{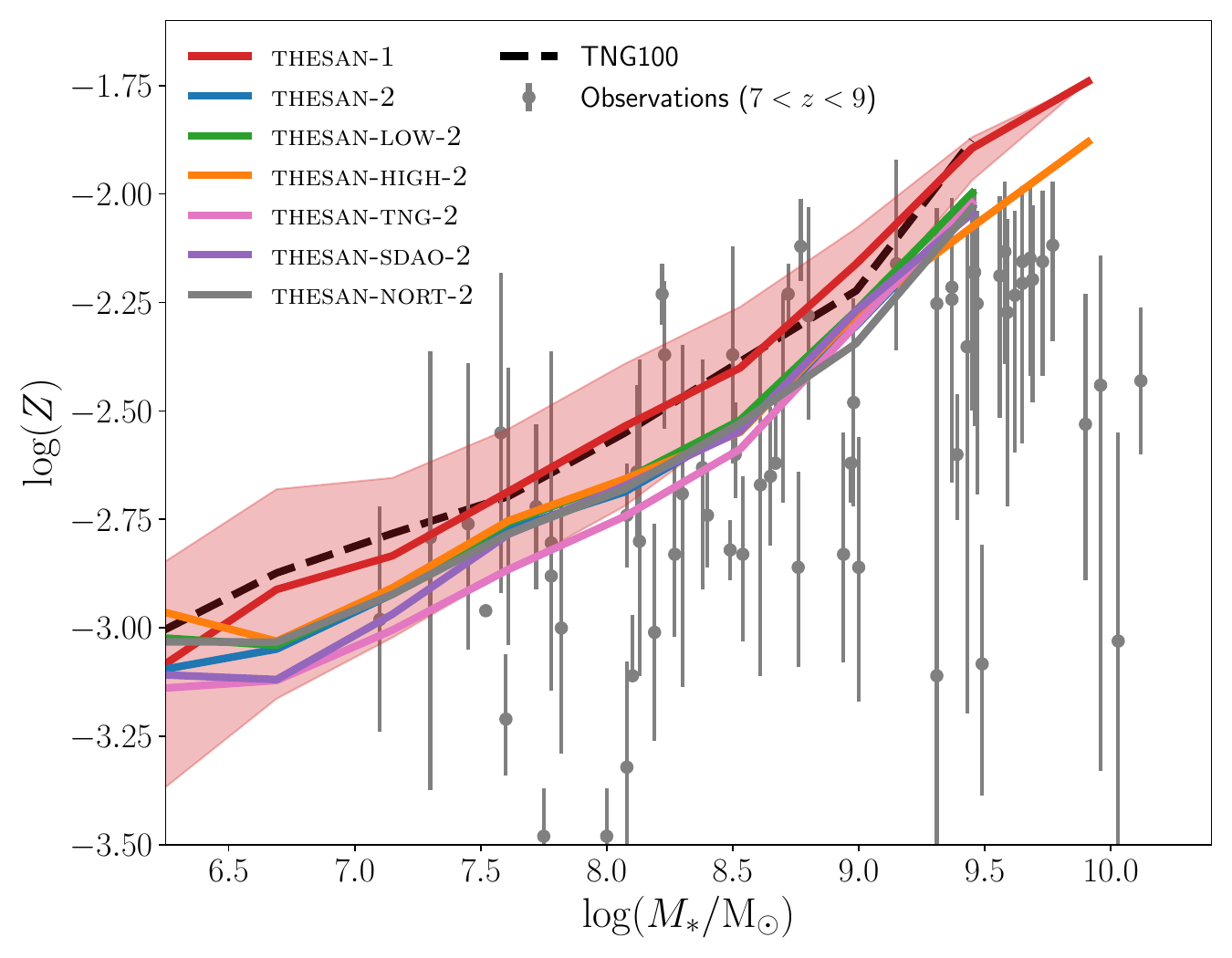}
    \caption{Same as Fig.~\ref{fig:MZR-z} but showing only the relation at $z=8$ for the different \thesan runs, as well as for TNG100 and for observations in the redshift range $7\leq z \leq9$.}
    \label{fig:MZR-all}
\end{figure}

The advent of \textit{JWST} has significantly advanced studies of the composition of galaxies deep in the reionization epoch beyond what can be inferred from their luminosities. In particular, the stellar mass--gas metallicity relation (MZR) has become an active area of interest. Fig.~\ref{fig:MZR-z} shows the MZR for \thesanone at five different redshifts, while the impact of different numerical and modelling assumptions is then shown (at $z=8$) in Fig.~\ref{fig:MZR-all}. In the latter, we also report the results computed from the Illustris-TNG 100 simulation. In both plots, the gas-phase O/H abundance ratio reported in observations has been converted to total gas-phase metallicities assuming a hydrogen mass fractions of 74 per cent and that oxygen makes up 35 per cent of the total metal mass. In summary, \thesanone shows no redshift evolution in the MZR, including its scatter (which, for the sake of clarity, we show only for $z=8$). At all redshifts investigated, \thesan appears to predict slightly too many metals at a fixed stellar mass compared to the bulk of available observations. This is in line with the likely underestimation of the cosmic dust content of galaxies (see \paperI), which leaves too many metals in the dust phase, thereby affecting the MZR. This particular aspect warrants further investigation, which we are carrying out. 

Finally, explored variations in the physical models in the \thesan simulations seem to leave little imprint on the MZR, as seen in Fig.~\ref{fig:MZR-all}. In fact, it appears that the main difference in the runs is due to the different resolutions employed, with \thesanone showing higher metallicities at any given stellar mass compared to the other runs. This is consistent with the slightly earlier onset and higher amount of star formation due to its higher resolution. However, \citet{ThesanHR} used the \thesanhr runs to show that the radiation modelling has an impact on the metal enrichment of the lowest-mass systems. Similarly, \citet{ThesanHR_noncdm} used the same simulations to show that some beyond-CDM models also suppress the metal production in low-mass systems. Thus, while resolution is a significant factor, other physical and numerical considerations can also play a role.

\subsection{UV slopes}
\dataused{merger trees, \Lya catalogs, SEDs}

The power-law index $\beta$ of the rest-frame galaxy UV continuum, often called UV slope, is influenced by the stellar ages and the metal content within the galaxy itself. As such, it is often used as an indicator of very young, extremely metal poor, dust free galaxies \citep[\eg][]{Cullen+2023}, which should approach the threshold value $\beta \approx -3$ \citep{Schaerer2002}. Additionally, the nebular emission from free-bound and two-photon processes can easily redden the slope, requiring a large ionizing escape fraction to attain $\beta \approx -3$, suppressing the contribution of nebular continuum. Therefore, a blue $\beta$ value is used to identify LyC leaker candidates and agents of reionization. Until the advent of the \textit{JWST}, the bluest galaxies known showed $\beta \approx -2.5$ \citep[e.g.][]{McLure+2018, Calzetti+1994}. Observational biases can also create artificially blue slopes \citep[\eg][]{Dunlop+2012, Rogers+2013}. Interestingly, the first galaxies robustly observed by the \textit{JWST} have $\beta$ values that are, on average, no bluer than in the local Universe \citep[\eg][]{Cullen+2023}.

\begin{figure}
    \centering
    \includegraphics[width=\columnwidth, trim={1px 2px 2px 2px},clip]{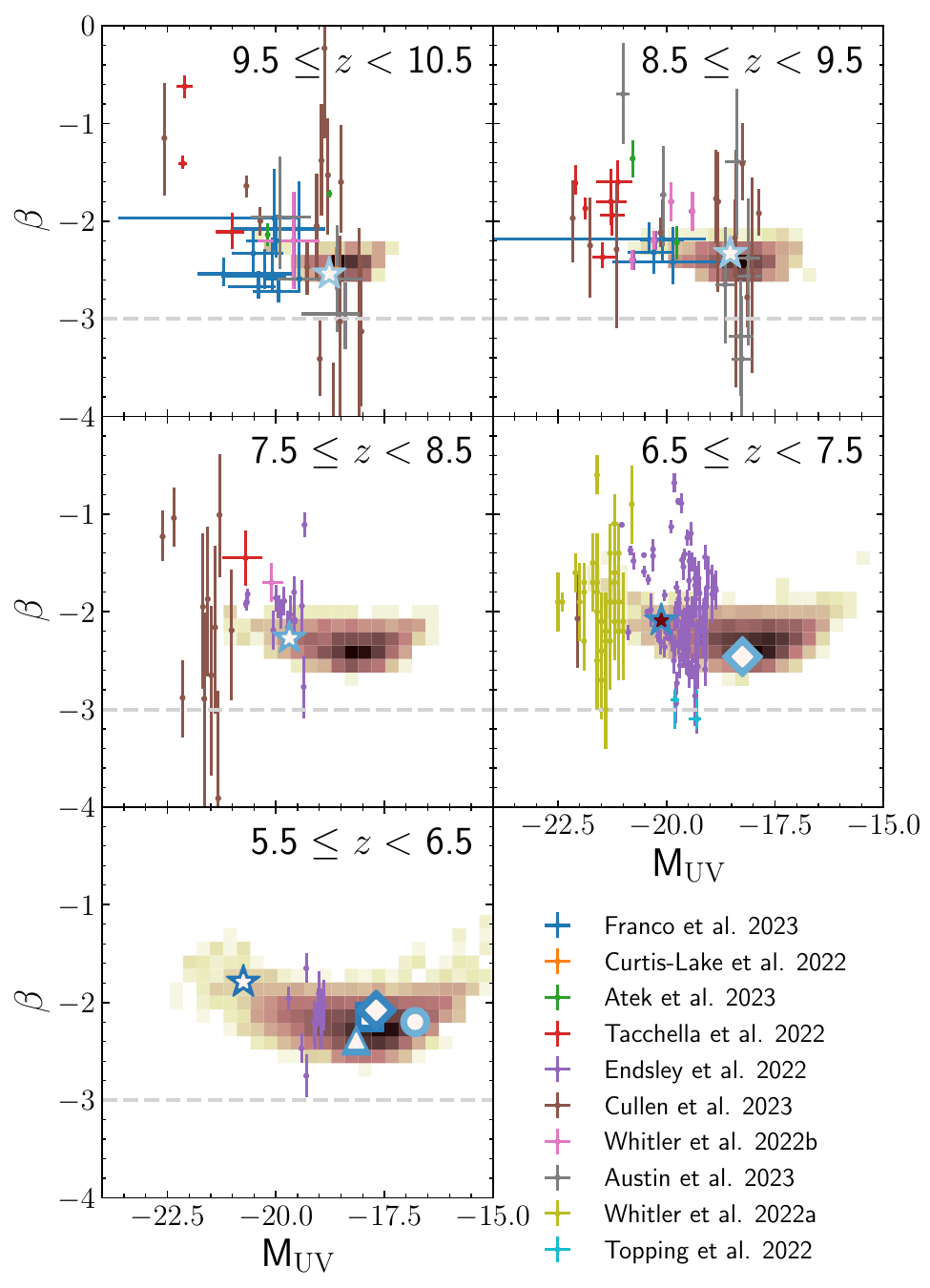}
    \caption{Distribution of UV magnitudes $M_\mathrm{UV}$ and slopes $\beta$ for candidate galaxies at $z=6$, $7$, $8$, $9$, $10$ in \thesanone (background shading) compared to recent observations from \citet{Atek+2023}, \citet{Austin+2023}, \citet{Cullen+2023}, \citet{Curtis-Lake+2023}, \citet{Endsley+2022}, \citet{Franco+2023}, \citet{Tacchella+2022}, \citet{Topping+2022} and \citet{Whitler+2023a,Whitler+2023b}. We show five randomly selected  simulated galaxies with different symbols, to highlight how they evolve in the diagram. The inner colour of the symbol contours encodes the galaxy escape fraction (with darker red pointing to a higher escape fraction), while the outer colour reflects the galaxy dust content (with darker blue indicating more dust). Finally, we indicate the threshold value $\beta = -3$ with a dashed grey horizontal line.}
    \label{fig:beta_muv}
\end{figure}

In Fig.~\ref{fig:beta_muv}, we show the $\beta$-slopes versus galaxy UV magnitude $M_\mathrm{UV}$ plane across five different redshift windows. The background two-dimensional histogram shows the predictions from \thesanone, generated using synthetic SEDs \citep[see Sec.~\ref{sec:sed_catalogs} and][for details on how they were obtained]{Thesan_lim}. The colour is logarithmically scaled from light to dark with the number of galaxies in the histogram bin. Overplotted with thin coloured symbols is a collection of recent \textit{JWST} observations\citep{Atek+2023, Austin+2023, Cullen+2023, Curtis-Lake+2023, Endsley+2022, Franco+2023}, \citet{Tacchella+2022, Topping+2022, Whitler+2023a,Whitler+2023b}. Additionally, to highlight the evolutionary paths of galaxies through this plane, we show with different thick symbols where five randomly-selected galaxies lie on this plane at each redshift. The symbol colours encode simulated properties of the galaxies, namely its escape fraction (inner colour, with darker red pointing to a higher escape fraction) and dust content (outer colour, with darker blue indicating more dust). Finally, the dashed grey horizontal line marks the theoretical threshold value of $\beta = -3$.

The \thesan galaxies lie in a U-shaped distribution (but see below). Objects with $M_\mathrm{UV} \approx -18$ tend to have the bluest UV slopes, averaging around $\beta \approx -2.5$. Brighter galaxies have, on average, larger dust masses which pushes $\beta$ to redder values. Conversely, UV-fainter galaxies typically have older stellar populations, which reduces their UV flux and increases their $\beta$ slopes. It is important to clarify that this U-shaped pattern does not imply that all UV-faint \thesan galaxies have redder slopes with respect to $M_\mathrm{UV} \approx -18$ ones. Rather, this is partially a consequence of the fact that synthetic SED (and therefore UV slopes) have been computed (to this date) only for the largest galaxies in the simulation, because of the relatively large computational time required. 
Using the $M_\mathrm{star}$--$M_\mathrm{UV}$ relation measured at $z=6$ by \citet{Bhatawdekar+2019}, we estimate that such a selection effect is significant for galaxies fainter than $M_\mathrm{UV} \gtrsim -18.3$. Brighter galaxies in \thesan generally align well with observed UV slopes, although observations exhibit a larger scatter compared to our simulations. However, it should be noted that most of the observed data points have large errors and are based on photometric measurements. Additionally, \thesanone covers only a relatively small volume, and is therefore unable to capture rare objects such as very UV-bright galaxies.

\subsection{Ionizing photon production efficiency}
\dataused{snapshots, offset files, \Lya catalogs, SEDs}

\begin{figure}
    \centering
    \includegraphics[width=\columnwidth]{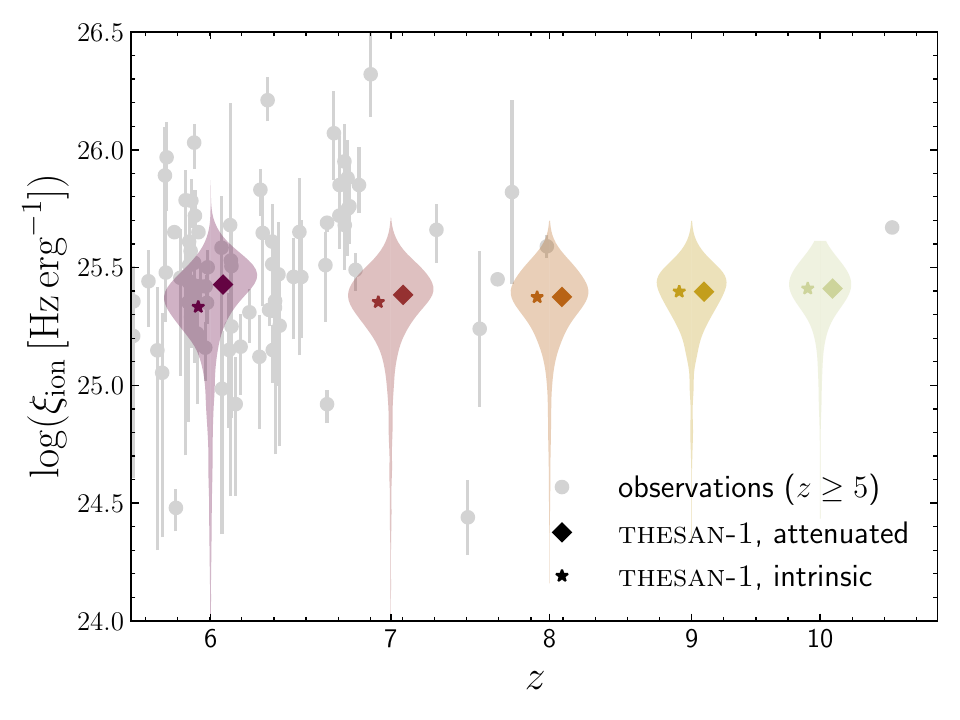}
    \caption{Redshift evolution of the ionizing photon production efficiency $\xi_\mathrm{ion}$ in \thesanone. The vertical violin plots show the distribution of values at a given redshift. For each of them, the right half shows the $\xi_\mathrm{ion}$ values including dust attenuation, while the left half shows the distribution of intrinsic values. The medians of these distributions are indicated by star and diamond symbols, respectively. The grey points correspond to a collection of observed values \citep{Bouwens+2016,Endsley&Stark2022,Bunker+2023,Jung+2023,Saxena+2023,Simmonds+2023}.}
    \label{fig:xi_ion_evol}
\end{figure}

A key quantity in our understanding of cosmic reionization as well as \highz galaxy formation is the ionizing photon production efficiency $\xi_\mathrm{ion}$. This is defined as the ratio between the rate of production of photons with energy $E_\gamma \geq 13.6\,eV$ and the UV luminosity of a galaxy, \ie $\xi_\mathrm{ion} = \dot{N}_\mathrm{ion} / M_\mathrm{UV}$, where $\dot{N}_\mathrm{ion}$ is the intrinsic rate of ionizing photons production and $M_\mathrm{UV}$ is the galaxy UV magnitude. Therefore, this value contains information about the stellar population of a galaxy (\eg age, IMF) and its dust obscuration (through the UV luminosity).

Thanks to the capabilities of the \textit{JWST}, this quantity can now be routinely measured in $z\gtrsim5$ galaxies. In Fig.~\ref{fig:xi_ion_evol}, we present these measurements \citep[from][]{Bouwens+2016,Endsley&Stark2022,Bunker+2023,Jung+2023,Saxena+2023,Simmonds+2023} as grey points, together with the predictions from the \thesanone simulation at integer redshifts. The latter are represented as violin plots, where the shaded areas indicate the distribution of $\xi_\mathrm{ion}$ values, and the median is marked by symbols. Specifically, the left half of each violin and the star symbols refer to the intrinsic values of $\xi_\mathrm{ion}$, while the right half of the distribution and the diamonds are computed including dust attenuation. The values of $\xi_\mathrm{ion}$ were computed combining the individual stellar particles of a galaxy with the BPASS v2.1 stellar library (the same used within the simulation) to determine the ionizing photons production rate, while the (dust-attenuated) UV luminosity is taken directly from the synthetic SED (see Sec.~\ref{sec:sed_catalogs}). The simulated values overlap well with the observed ones, but with the caveat that \thesanone does not contain galaxies with $\xi_\mathrm{ion} \gtrsim 25.8$ that are however found in observations. This discrepancy might be due to the limited simulated volume, which fails to capture rare objects, or to the assumed stellar population model and IMF. 

We do not observe any evolution with time in the bulk of the distributions, nor significant dependence on UV magnitude or stellar mass (not shown in the figure). At the same time, the $\xi_\mathrm{ion}$ of an individual galaxy varies strongly during its lifetime. Taken together, these results indicate that the population-averaged ionizing photon output does not vary significantly over the interval $6 \leq z \leq 10$. Instead, the growth of the galaxy population and the increase of the total stellar mass in the Universe is what advances and accelerates reionization towards $z\sim6$. It should be noted, however, that the high-$\xi_\mathrm{ion}$ tail of the distribution is cut short at higher redshift. This truncation is due to a combination of poorer statistics (\ie~fewer galaxies) and a general absence of massive, dusty galaxies. The latter, in fact, lowers the observed UV luminosity, increasing $\xi_\mathrm{ion}$ at a fixed production rate of ionizing photons. Therefore, the formation of an increasing amount of dust can drive up $\xi_\mathrm{ion}$ values at lower redshifts, even if the intrinsic properties of galaxies remain unchanged. This can be appreciated comparing the two sides of the violin plots, which makes apparent that dust obscuration starts to have an impact on the distribution of observed $\xi_\mathrm{ion}$ only at $z\lesssim7$.

\section{Summary and Conclusions}
\label{sec:conclusions}
As part of this paper, we have publicly released the entire set of data and data products generated from the \thesan simulation suite, including the raw simulation outputs. These resources are freely accessible to the research community and come with complete documentation. They can be downloaded from the project's official website: \url{https://thesan-project.com}. This release aims to facilitate further research and progress in the interconnected fields of galaxy formation and cosmic reionization.

\thesan is a suite of radiation-magneto-hydrodynamical simulations designed to simultaneously capture the cosmological scales required for reionization modelling and the $\sim 100$\,pc resolved physics responsible for galaxy formation in the high-redshift ($z\geq5$) Universe. It incorporates the IllustrisTNG galaxy formation model, self-consistent radiation transport of photons emitted by both stars (including binary systems) and quasars, a sophisticated model for the creation, evolution, and destruction of cosmic dust, and variance-suppressed initial conditions. The main series of simulations follows a large volume (95.5\,cMpc)$^3$ of the Universe under different physical models for the escape of ionizing radiation from galaxies and for the nature of dark matter. Alongside these, we have also released data from the \thesanhr series, which consists of even higher-resolution simulations in smaller volumes, specifically designed to investigate non-trivial effects of reionization on the properties of lower-mass galaxies.

\thesan augments traditional simulation outputs, such as snapshots and group catalogs, with novel high-time-cadence Cartesian outputs and a wide range of post-processing data products. These are thoroughly described in this paper with additional details in the online documentation, and are publicly available as well. In order to ease the usage of \thesan by researchers with varying expertise, we have integrated support for \thesan data into widely-used analysis tools.

The \thesan simulations offers robust predictions that align well with recent observations from the \textit{JWST} concerning galaxies in the first billion years of the Universe, as we have shown in the second part of this manuscript. Specifically, \thesan matches key relations such as the galaxy main sequence, the mass--metallicity relation, and the UV slope--stellar mass relation at $z\geq6$. Additionally, the simulations provide forecasts for the fraction of dust-obscured star formation as a function of galaxy mass. Leveraging the physical variations simulated within the \thesan project, we show how these important quantities are influenced by the still-uncertain physics at high redshifts.

\thesan represents the present state of the art in simulating the first billion years of cosmic history, from IGM to CGM and galactic scales. We have made the entire data of the project fully available to the community in the hope that researchers across the world will contribute to fully exploiting the potential of these simulations. As new data are generated and vetted, they will be added to the \thesan repository and accompanied by updated documentation on the project website.

Finally, the \thesan project is an ongoing and continually evolving initiative. Broadly speaking, we are currently focusing on three parallel programmes embodying the near-term futures of \thesan. First, we aim to increase the simulated volume to both capture rarer objects and reach sufficiently-large scales to more effectively interface with 21\,cm cosmology, including improved sampling of lower modes of the power spectrum. Second, we are entering the production stage of developing zoom-in re-simulations that incorporate a model for the multi-phase ISM, the large-scale radiation field of \thesan, and extend to significantly lower redshifts than the original \thesan volume. Third, we plan to conduct re-simulations of individual sub-volumes within the \thesan box, each with different reionization histories and varying environmental overdensities, to rigorously assess the impact of local reionization on a statistically-significant population of highly-resolved galaxies. In summary, the \thesan project invites extensive direct and indirect community collaboration to advance these targeted efforts, thereby promising to transform our understanding of cosmic reionization and early galaxy formation.

\section*{Acknowledgments}
\rev{}{We are grateful to the referee, Nick Gnedin, for his insightful comments, which improved our manuscript.} 
EG acknowledges support from the CANON Foundation Europe through the Canon Fellowship program during part of the work presented in this paper. 
RK and AS acknowledge support under Institute for Theory and Computation Fellowships at the Center for Astrophysics $\vert$ Harvard \& Smithsonian, and AS for Program number \textit{HST}-HF2-51421.001-A provided by NASA through a grant from the Space Telescope Science Institute, which is operated by the Association of Universities for Research in Astronomy, incorporated, under NASA contract NAS5-26555.
MV acknowledges support through NASA ATP grants 16-ATP16-0167, 19-ATP19-0019, 19-ATP19-0020, 19-ATP19-0167, and NSF grants AST-1814053, AST-1814259,  AST-1909831 and AST-2007355. 
The authors gratefully acknowledge the Max Planck Computing and Data Facility (\url{https://www.mpcdf.mpg.de/}) for support in hosting data and releasing them to the public, as well as the Gauss Centre for Supercomputing e.V. (\url{www.gauss-centre.eu}) for funding this project by providing computing time on the GCS Supercomputer SuperMUC-NG at Leibniz Supercomputing Centre (\url{www.lrz.de}). Additional computing resources were provided by the Extreme Science and Engineering Discovery Environment (XSEDE), at Stampede2 through allocation TG-AST200007, by the NASA High-End Computing (HEC) Program through the NASA Advanced Supercomputing (NAS) Division at Ames Research Center, and the Engaging cluster
supported by the Massachusetts Institute of Technology.
We are thankful to the community developing and maintaining software packages extensively used in our work, namely: \texttt{matplotlib} \citep{matplotlib}, \texttt{numpy} \citep{numpy}, \texttt{scipy} \citep{scipy}, \texttt{cmasher} \citep{cmasher} and \texttt{CoReCon} \citep{corecon}.

\section*{Data availability}
All data produced within the \thesan project are made fully and openly available at \url{https://thesan-project.com}, including extensive documentation and usage examples. We invite inquiries and collaboration requests from the community.

\bibliographystyle{mnras}
\bibliography{bibliography}

\bsp 
\label{lastpage}
\end{document}